\documentclass[letterpaper,12pt]{JHEP3}
\pdfoutput=1
\usepackage{amsmath}
\usepackage{amssymb}
\usepackage{epsfig}
\newcommand{\labell}[1]{\label{#1}}
\def\({\left(} \def\){\right)}
\def\[{\left[} \def\]{\right]}
\def\al{\alpha} \def\bt{\beta}
\def\del{{\partial}}
\def\hz{\mathrel{\mathop h^{\scriptscriptstyle{(0)}}}{}\!\!}
\def\Xz{\mathrel{\mathop X^{\scriptscriptstyle{(0)}}}{}\!\!}
\def\Xo{\mathrel{\mathop X^{\scriptscriptstyle{(1)}}}{}\!\!}
\def\gz{\mathrel{\mathop g^{\scriptscriptstyle{(0)}}}{}\!}
\def\go{\mathrel{\mathop g^{\scriptscriptstyle{(1)}}}{}\!}
\def\gs{\mathrel{\mathop g^{\scriptscriptstyle{(2)}}}{}\!}

\def\hz{\mathrel{\mathop h^{\scriptscriptstyle{(0)}}}{}\!}
\def\ho{\mathrel{\mathop h^{\scriptscriptstyle{(1)}}}{}\!}

\newcommand{\non}{\nonumber \\}

\newcommand{\be}{\begin{equation}}
\newcommand{\ee}{\end{equation}}
\newcommand{\bea}{\begin{eqnarray}}
\newcommand{\eea}{\end{eqnarray}}
\newcommand{\ba}{\begin{eqnarray}}
\newcommand{\ea}{\end{eqnarray}}

\newcommand{\beq}{\begin{equation}}
\newcommand{\eeq}{\end{equation}}
\newcommand{\beqa}{\begin{eqnarray}}
\newcommand{\eeqa}{\end{eqnarray}}
\newcommand{\beqar}{\begin{eqnarray*}}
\newcommand{\eeqar}{\end{eqnarray*}}

\newcommand{\reef}[1]{(\ref{#1})}
\newcommand{\ssc}{\scriptscriptstyle}
\newcommand{\eg}{{\it e.g.,}\ }
\newcommand{\ie}{{\it i.e.,}\ }

\newcommand{\mt}[1]{\textrm{\tiny #1}}

\newcommand{\K}{\mathcal{K}}
\newcommand{\cL}{\mathcal{L}}
\newcommand{\M}{\mathcal{M}}

\newcommand{\R}{\mathcal{R}}

\newcommand{\tL}{\tilde{L}}

\newcommand{\veps}{\varepsilon}
\newcommand{\eps}{\epsilon}
\newcommand{\la}{\lambda}
\def\al{\alpha}
\def\bt{\beta}
\def\ga{\gamma}
\def\de{\delta}
\def\del{{\partial}}
\newcommand{\lp}{\ell_{\mt P}}

\newcommand{\fin}{f_\infty}

\newcommand{\hi}{{\hat \imath}}
\newcommand{\hj}{{\hat \jmath}}

\newcommand{\ads}{a_d^*}
\renewcommand{\r}[2]{R_\mt{#1}^{\,#2}}

\preprint{arXiv:1101.5813 [hep-th]}

\title{On Holographic Entanglement Entropy\\ and Higher Curvature Gravity}

\author{Ling-Yan Hung, Robert C. Myers and Michael Smolkin\\
{\it Perimeter Institute for Theoretical Physics,\\ 31 Caroline Street North, Waterloo,
Ontario N2L 2Y5, Canada}}

\abstract{We examine holographic entanglement entropy with higher
curvature gravity in the bulk. We show that in general Wald's formula
for horizon entropy does not yield the correct entanglement entropy.
However, for Lovelock gravity, there is an alternate prescription which
involves only the intrinsic curvature of the bulk surface. We verify that
this prescription correctly reproduces the universal contribution to
the entanglement entropy for CFT's in four and six dimensions. We also
make further comments on gravitational theories with more general higher
curvature interactions.}

\begin{document}

\section{Introduction}

In their seminal work \cite{rt1}, Ryu and Takayanagi made a proposal
for the calculation of entanglement entropy of the boundary field
theory in the context of gauge/gravity duality --- see also
\cite{rt2,rt3}. Their approach is both simple and elegant. Given a
particular spatial region $V$ in the boundary theory, their proposal
for the entanglement entropy between this region and its complement is
 \be
S(V) = \frac{2\pi}{\lp^{d-1}}\ \mathrel{\mathop {\rm
ext}_{\scriptscriptstyle{m\sim V}} {}\!\!} \left[A(m)\right]
 \labell{define}
 \ee
where $m\sim V$ indicates that $m$ is a bulk surface that is homologous
to the boundary region $V$ \cite{head,furry}. The symbol `ext'
indicates that one should extremize the area over all such surfaces
$m$.\footnote{If the calculation is done in a Minkowski signature
background, the extremal area is only a saddle point. However, if one
first Wick rotates to Euclidean signature, the extremal surface will
yield the minimal area. In either case, the area must be suitably
regulated to produce a finite answer. Note that for a $d$-dimensional
boundary theory, the bulk has $d+1$ dimensions while the surface $m$
has $d-1$ dimensions. We are using `area' to denote the
($d-1$)-dimensional volume of $m$.} This result applies where the bulk
is described by classical Einstein gravity. Hence we might note the
similarity between this expression \reef{define} and that for black
hole entropy. In a more general holographic framework, one would
evaluate the area using the Einstein-frame metric \cite{head}.

There is also a fair amount of evidence to support this conjecture
\cite{head}:
\begin{itemize}
\item As shown in \cite{rt1}, it reproduces precisely the
    entanglement entropy of a $d=2$ CFT for an interval of length
    $\ell$ on a circle of circumference $2\pi R$ \cite{cardy0,finn}
  \be
  S(V)=\frac{c}{3}\log\left(\frac{2R}{\delta}\,\sin\frac{\ell}{2R}
  \right)\,, \labell{twod} \ee
    where $c$ is the central charge and $\delta$ is a
    short-distance cut-off. While this result applies for the
    vacuum, this holographic expression \reef{define} can easily be
    shown to reproduce the expected entanglement entropy at finite
    temperature for $d=2$.
\item The leading (divergent) term in $S(V)$ takes precisely the
    form expected for the `area law' contribution to the
    entanglement entropy in a $d$-dimensional CFT \cite{rt2,rt3}.
    That is, the leading contribution is proportional to
    $A(\partial V)/\delta^{d-2}$.
\item As expected, if one considers a pure state in the boundary
    CFT (which is dual to a fixed bulk geometry without a horizon),
    one finds that $S(V)=S(\bar V)$ where $\bar V$ denotes the
    complement of $V$.
\item Given two boundary regions, $V$ and $U$, one readily shows
    that this construction \reef{define} obeys the necessary
    inequality known as `strong subadditivity' \cite{head2}. That
    is,
 \be S(V\cup U) + S(V\cap U)\le S(V) + S(U)\,. \labell{ssub} \ee
\item In a slightly different context, this approach reproduces the
    Bekenstein-Hawking entropy of an eternal black hole. Recall
    that in the context of the AdS/CFT correspondence, the two
    asymptotic boundaries of an eternal black hole are associated
    with the original CFT and its thermofield double \cite{juan}.
    The horizon entropy can then be associated with the
    entanglement entropy between these two sets of degrees of
    freedom. Applying eq.~\reef{define} in this context, the region
    $V$ becomes the entire boundary (of one asymptotic region) and
    $m$ is then the black hole horizon \cite{head}.
\end{itemize}

A standard approach to calculating entanglement entropy (EE) makes use
of the replica trick \cite{cardy0,callan}. Unfortunately, this
technique makes use of a singular background geometry as an
intermediate tool and the natural holographic translation involves a
singular bulk manifold \cite{furry}. It seems that without a full
understanding of string theory or quantum gravity in the bulk, we will
not be able to work with this bulk geometry in a controlled way. In
particular, it is not possible to properly evaluate the saddle-point
action in the gravitational bulk theory. Hence despite various efforts
\cite{furry}, a constructive proof of the Ryu-Takayanagi proposal
\reef{define} is still unknown.\footnote{However, see \cite{circle4}
for recent progress in this direction.} With a complete derivation, one
could easily take into account the appearance of higher curvature terms
in the bulk gravity theory, \eg to calculate finite $N_c$ or finite
$\lambda$ corrections to holographic EE. Without such a derivation in
hand, we set out here to explore holographic entanglement entropy in
higher curvature gravity.

Previous results provide important suggestions as to how we should
proceed to extend eq.~\reef{define} in the presence of higher curvature
interactions in the bulk theory. First of all, as long as the
prescription is one of minimizing a `surface functional', we expect to
have a formalism where the holographic EE satisfies subadditivity
\reef{ssub}, as in \cite{head2}. Hence the question becomes how to
define the appropriate surface functional given a particular higher
curvature gravity action. As noted above, there is a close connection
between holographic entanglement entropy and black hole entropy. In
particular, to extend the description of the horizon entropy of an
eternal black hole in terms of holographic EE, it must be the case that
evaluating the new surface functional on an event horizon yields the
correct black hole entropy in the higher curvature gravity theory. A
first suggestion then would be that the surface functional simply
coincides with the expression for Wald's formula \cite{wald1} for black
hole entropy in higher curvature gravity. Unfortunately, as we will
show below, this proposal fails! However, we must still demand that the
new functional should be compatible with Wald's formula on an event
horizon. To make further progress, our discussion will focus on
Lovelock gravity \cite{lovel} in the bulk. We regard the latter as
simply a convenient toy model with which we can easily make explicit
calculations and one which may provide some useful insights into more
general bulk theories.

An overview of the paper is as follows: We begin with a brief review of
of some useful background material in section \ref{esse}. We describe
Lovelock gravity and Wald's entropy formula. We also describe another
prescription, which we denote $S_\mt{JM}$, for black hole entropy
specifically derived for Lovelock gravity \cite{ted1}. Further we also
review a field theory calculation of the universal contribution to
entanglement entropy in even dimensional CFT's \cite{rt2,solo}. In
section \ref{not}, we show that the suggestion of replacing
eq.~\reef{define} by an extremization of Wald's entropy formula fails
to provide the correct EE in general. For Lovelock gravity, this leaves
us with the $S_\mt{JM}$ prescription and we verify this proposal by
comparing the universal contribution to the holographic EE to the
analogous CFT results for a variety of geometries in four and six
dimensions in sections \ref{EE4} and \ref{EE6}. We return to
considering holographic EE for general gravitational actions in section
\ref{general}. In particular, our analysis there points out a new
ambiguity in the prescription for holographic EE in Lovelock gravity.
However, we are able to eliminate this potential ambiguity by
considering the details of the variational problem. We conclude with a
brief discussion of our results, including some interesting
applications, in section \ref{discuss}. There are also four appendices
which provide some of the useful technical details.

While proceeding with this project, we learned that the same topic was
also being studied by J.~de Boer, M.~Kulaxizi and A.~Parnachev  --- see
talk by M.~Kulaxizi \cite{friends1}. Their results appear in
\cite{friends2}. We also note that the effect on holographic
entanglement entropy from a certain higher curvature interaction, the
gravitational Chern-Simons term, in three-dimensional AdS space was
studied by \cite{sun}.

\section{A few preliminaries} \label{esse}

Our primary aim in this paper is to explore the contribution of higher
curvature interactions in the bulk gravity theory to holographic
entanglement entropy. In the next few sections, we will focus our
attention on Lovelock gravity \cite{lovel}. The latter provides a
useful toy model where one can readily perform explicit calculations.
We return to more general considerations in sections \ref{general}.
Hence, we begin below with a brief review of Lovelock gravity to set
the context for our discussion in the following sections. Next, as
alluded to above, there is a close connection between black hole
entropy and holographic entanglement entropy and so we also review two
proposals for the latter in higher curvature gravity. First, there is
Wald's entropy formula \cite{wald1}, which can be applied for any
covariant gravity action, and then an earlier result derived
specifically for Lovelock gravity \cite{ted1}. Finally, setting aside
gravity and holography, we also review a calculation of entanglement
entropy in even dimensional CFT's \cite{rt2,solo}. In these purely
field theoretic calculations, the universal contribution to the
entanglement entropy is related to the central charges in the trace
anomaly. The results for general CFT's must be reproduced in our
calculations of the holographic entanglement entropy and so provides a
crucial test in extending the latter to higher curvature gravity.

\subsection{Lovelock gravity} \label{glove}

Lovelock gravity \cite{lovel} is the gravitational theory in higher
dimensions with higher curvature interactions proportional to the Euler
density of higher even dimensional manifolds. The general Lovelock
action in $d+1$ dimensions can be written as\footnote{Here, we follow
closely the notation of \cite{jan3}.}
 \beq
I = \frac{1}{2\lp^{d-1}} \int \mathrm{d}^{d+1}x \, \sqrt{-g}\, \left[
\frac{d(d-1)}{L^2} + R +
\sum_{p=2}^{\left\lfloor\frac{d+1}{2}\right\rfloor}
c_p\,L^{2p-2}\,\cL_{2p}(R) \right]\,,
 \labell{GBactg}
 \eeq
where $\left\lfloor\frac{d+1}{2}\right\rfloor$ denotes the integer part
of $(d+1)/2$ and $c_p$ are dimensionless coupling constants for the
higher curvature terms. These higher order interactions are defined as
 \beq
\cL_{2p}(R) \equiv \frac{1}{2^p}\ \delta_{\mu_1\,\mu_2\,\cdots\,
\mu_{2p-1}\,\mu_{2p}}^{\nu_1\,\nu_2\,\cdots\, \nu_{2p-1}\,\nu_{2p}}\
R^{\mu_1\mu_2}{}_{\nu_1\nu_2}\,\cdots\,
R^{\mu_{2p-1}\mu_{2p}}{}_{\nu_{2p-1}\nu_{2p}}\,,
 \labell{term0}
 \eeq
which is proportional to the Euler density on a 2$p$-dimensional
manifold. Here, we are using $\delta_{\mu_1\,\mu_2\,\cdots\,
\mu_{2p-1}\,\mu_{2p}}^{\nu_1\,\nu_2\,\cdots\, \nu_{2p-1}\,\nu_{2p}}$ to
denote the totally antisymmetric product of $2p$ Kronecker delta
symbols. Of course, the cosmological constant and Einstein terms could
be incorporated into this scheme as $\cL_0$ and $\cL_1$, respectively.
However, we exhibit them explicitly above to establish our
normalization for the Planck length, as well as the length scale $L$.
By construction, it is clear that in $d+1$ dimensions, all Lovelock
$\cL_p$ terms with $p> (d+1)/2$ must vanish --- hence the explicit
restriction on the sum in eq.~\reef{GBactg} is not really required. For
$p=(d+1)/2$, $\cL_{2p}$ is topological. While this last term does not
contribute to the gravitational equations of motions, it can contribute
to black hole entropy \cite{ted1,cosmic}.

The original motivation to construct this action \reef{GBactg} was that
the resulting equations of motion are only second order in derivatives
\cite{lovel}. Another interesting feature of these theories is the
equivalence between metric and Palatini formulations \cite{pala}.
Earlier studies also found exact (asymptotically flat) black hole
solutions to the classical equation of motion \cite{GBbh} and the exact
form of the Gibbons-Hawking surface term is known \cite{surf}.
Recently, there has been renewed interest in these theories in the
context of the AdS/CFT correspondence. In particular, asymptotically
AdS black hole solutions were found for Lovelock gravity
\cite{jan3,fate,GBadsbh}. These exact solutions then proved useful in
discussions of holographic hydrodynamics and consistency of the
boundary CFT \cite{jan3,fate,EtasGB,jan2}. Further these models have
also been shown to satisfy a holographic c-theorem \cite{cosmic,jtliu}.

Anticipating our application to the AdS/CFT correspondence, we have
explicitly included a negative cosmological constant in the action
\reef{GBactg}. The theory then has AdS$_{d+1}$ vacua with a curvature
scale $\tL^2=L^2/\fin$ where $\fin$ is a root of the following
expression:
\be 1=\fin-\sum_{p=2}^{\left\lfloor d/2\right\rfloor}\lambda_p\,
\left(\fin\right)_{\vphantom{Z}}^{p}\,. \labell{roots} \ee
To simplify this expression, we have introduced the following notation
 \be
\lambda_p=(-)^p\,\frac{(d-2)!}{(d-2p)!}\ c_p\,.
 \labell{grunt}
 \ee
Note that the topological term (\ie $p=(d+1)/2$) does not contribute to
determining the AdS scale and so the upper limit on the sum here is not
the same as in the action \reef{GBactg}. In general, this equation
yields $\left\lfloor d/2\right\rfloor$ different roots for $\fin$. We
are only interested in the positive real roots, since these correspond
to AdS$_{d+1}$ vacua. However, for many of these roots, the graviton is
in fact a ghost-like excitation, \ie its kinetic term has the wrong
sign \cite{GBghost,old1} and further, even if the latter problem is
evaded, the vacuum typically does not support nonsingular black hole
solutions \cite{old1}. In fact, there is at most one root which
yields a ghost-free AdS vacuum which supports black hole solutions, as
described in detail in \cite{old1}. Further, in a regime where the
$\lambda_p$ are not large, this will be the smallest positive root and
it is continuously connected to the single root ($\fin=1$) that remains
in the Einstein gravity limit, \ie $\lambda_p\rightarrow 0$.
Implicitly, we will be working in this regime of the coupling space and
with this particular root in the following.

Of particular interest in the following, will be the central charges of
the boundary CFT for even $d$.  For any CFT in an even number of
dimensions, the central charges can be defined in terms of the trace
anomaly --- see eq.~\reef{trace} and the discussion in section
\ref{EEcft}. Now in the context of the AdS/CFT correspondence, general
techniques have been developed to holographically evaluate the trace
anomaly and determine the corresponding central charges \cite{sken}.
When the bulk theory is described by Einstein gravity, one finds that
all of the charges are essentially equal, being determined by the ratio
$(\tL/\lp)^{d-1}$. However, with the introduction of higher curvature
terms in the bulk gravity, the central charges become functions of the
new (dimensionless) couplings, as well as the ratio of the AdS scale to
the Planck scale, and so the charges can be (at least partially)
distinguished in such an extended holographic set up
\cite{highc,old2,jan3,jan2}.

In general, determining all of the central charges is a fairly involved
calculation, however, there is a simple short-cut to calculate $A$
presented in \cite{adam}. Given any general covariant action for the
bulk gravity theory, $A$ is determined by simply evaluating the value
of the Lagrangian in the AdS$_{d+1}$ vacuum. With the conventions of
\cite{cosmic,cthem}, which we have adopted here,
 \be
A=-\frac{\pi^{d/2}\,\tilde L^{d+1}}{d\,\Gamma\left(d/2 \right)}
\left.{\cal L}\right|_{AdS}\,.
 \labell{adamA}
 \ee
We emphasize that the right-hand side is evaluated with the theory in
Minkowski signature and we refer the interested reader to \cite{cosmic}
for further details. In the case of the Lovelock action \reef{GBactg},
evaluating the above expression is a straightforward exercise, which
yields
 \be
A=\frac{\pi^{d/2}}{\Gamma\left(d/2
\right)}\,\frac{\tL^{d-1}}{\lp^{d-1}}\left(1-\sum_{p=2}^{\left\lfloor
d/2\right\rfloor}\frac{(d-1)\,p}{d+1-2p}\,\lambda_p\,
\left(\fin\right)_{\vphantom{Z}}^{p-1}\right)
 \labell{llA}
\ee
Here we have used eqs.~\reef{roots} and \reef{grunt} to arrive at this
result. Note that in the case of the topological term with $p=(d+1)/2$,
one would add an extra term to eq.~\reef{llA} of the form
 \be
\delta A=\frac{\pi^{d/2}}{\Gamma\left(d/2
\right)}\,\frac{L^{d-1}}{\lp^{d-1}}\times
(-)^{\frac{d+3}{2}}\frac{(d+1)!}{2d} c_{\frac{d+1}{2}}\,.
 \labell{llAtop}
 \ee

\subsection{Horizon entropy} \label{wald}

As noted in the introduction, there is a close connection between black
hole entropy and holographic entanglement entropy. For any (covariant)
theory of gravity, the black hole entropy can be calculated using
Wald's entropy formula \cite{wald1}
 \beq
S = -2 \pi \int_\mt{horizon} d^{d-1}x\sqrt{h}\
\frac{\partial{\cL}}{\partial R^{\mu\nu}{}_{\rho\sigma}}\,
\hat{\veps}^{\,\mu\nu}\,\hat{\veps}_{\rho\sigma}\,,
 \labell{Waldformula}
 \eeq
where $\mathcal{L}$ denotes the gravitational Lagrangian and
$\hat{\veps}_{\mu\nu}$ is the binormal to the horizon normalized by
$\hat{\veps}_{\mu\nu}\hat{\veps}^{\mu\nu}=-2$ (assuming a Minkowski
signature), while $h$ is the determinant of the induced metric
$h_{\al\bt}$ on the horizon. Now this prescription can easily be
applied to the Lovelock theory \reef{GBactg} yielding
 \beq
S_\mt{W} = \frac{2 \pi}{\lp^{d-1}} \int_\mt{horizon}
d^{d-1}x\,\sqrt{h}\,\left[ 1+\sum_{p=2}^{\left\lfloor\frac{d+1}{2}
\right\rfloor} p\,c_p\,L^{2p-2}\,\cL_{2p-2}(R^\parallel) \right]\,.
 \labell{Waldformula2}
 \eeq
Here $R^\parallel$ denotes the components of the curvature tensor
projected onto the horizon, \ie
 \be
\left[R^\parallel\right]{}^{\alpha\beta}{}_{\gamma\delta}=
h^\alpha{}_{\alpha'}\, h^\bt{}_{\bt'} \,h_\ga{}^{\ga'}
\,h_\de{}^{\de'}\,R^{\al' \bt'}{}_{\ga' \de'}\,.
 \labell{paral}
 \ee

We note, however, that this expression for the horizon entropy is not
unique. In particular, black hole entropy in the Lovelock theory was
studied in \cite{ted1}, which preceded (and in part, motivated) the
derivation of Wald's formula \reef{Waldformula}. Using a Hamiltonian
approach, this earlier work \cite{ted1} derived the following
expression
 \beq
S_\mt{JM} = \frac{2 \pi}{\lp^{d-1}} \int_\mt{horizon}
d^{d-1}x\,\sqrt{h}\,\left[
1+\sum_{p=2}^{\left\lfloor\frac{d+1}{2}\right\rfloor}
p\,c_p\,L^{2p-2}\,\cL_{2p-2}(\R) \right]\,,
 \labell{Waldformula3}
 \eeq
where $\R^{\al\bt}{}_{\ga\de}$ are the components of the intrinsic
curvature tensor of the slice of the event horizon on which this
expression is evaluated.

In fact, there is no disagreement between eqs.~\reef{Waldformula2} and
\reef{Waldformula3} in the context for which they were derived. Both
derivations \cite{wald1,ted1} assumed that the relevant horizon was a
Killing horizon, \ie the black hole background is stationary with a
Killing vector $\chi^{\mu}$ which becomes null on the horizon. The
geometry is remarkably constrained in this case \cite{rack} and it is
straightforward to show, in particular on the bifurcation surface, that
the extrinsic curvatures vanishes. Now recall that the Gauss-Codazzi
equations relate the intrinsic curvature to the projection of the full
spacetime curvature with \cite{mtw}
 \be
\left[R^\parallel\right]{}_{\al\bt\ga\de}=\R_{\al\bt\ga\de}-\sum_{i=1}^2\,
\eta_{\hi \hj}\, \left(K^{\hi }_{\al\ga}K^{\hj}_{\bt\de}-K^{\hi
}_{\al\de} K^{\hj}_{\bt\ga}\right)\,.
 \labell{gacod}
 \ee

To describe this result, we need to introduce some formalism, which
will also be useful in later discussion.\footnote{The surface of
interest in the present discussion is a slice of the black hole
horizon, however, we will also apply the same formalism to the bulk
surface used in calculating holographic entanglement entropy. Both are
codimension two surfaces embedded in the relevant spacetime. Let us
also comment on our index conventions throughout the paper. Directions
in the full (AdS) geometry are labeled with letters from the second
half of the Greek alphabet, \ie $\mu,\nu,\rho,\cdots$. Letters from the
`second' half of the Latin alphabet, \ie $i,j,k,\cdots$, correspond to
directions in the background geometry of the boundary CFT. Meanwhile,
directions along the entangling surface in the boundary are denoted
with letters from the beginning of the Latin alphabet, \ie
$a,b,c,\cdots$, and directions along the corresponding bulk surface are
denoted with letters from the beginning of the Greek alphabet, \ie
$\al,\bt,\ga,\cdots$. Finally, we use hatted letters from the later
part of the Latin alphabet to denote the frame or tangent indices in
the transverse space to both of these surfaces, \ie $\hi,\hj$.
\label{footy}} There is a pair of unit vectors $n^\hi_\mu$ (with
$\hi=1,2$) which are orthogonal to the surface (on which
eq.~\reef{gacod} is evaluated) and to each other. Then $\eta^{\hi
\hj}=n^{\hi}_\mu n^{\hj\,\mu}$ is the Minkowski\footnote{If the
embedding geometry had a Euclidean signature, then this transverse
metric would simply be a Kronecker delta $\delta^{\hi\hj}$.} metric in
the transverse tangent space spanned by these vectors and $\eta_{\hi
\hj}$ is the inverse of this metric. We also have tangent vectors
$t_\al^\mu$ along the surface, which are defined in the usual way with
$t_\al^\mu=\partial X^\mu/\partial\sigma^\al$ where $X^\mu$ and
$\sigma^\al$ are the coordinates in the full embedding space and along
the surface, respectively. The induced metric is then given by
$h_{\al\bt}=t^\mu_\al\, t^\nu_\bt\, g_{\mu\nu}$. We may also define
this induced metric as a bulk tensor with
$h_{\mu\nu}=g_{\mu\nu}-\eta_{\hi\hj}\,n^\hi_\mu n^\hj_\nu$. The second
fundamental forms are defined for the surface with $K^{\hi}_{\al\bt} =
-t^\mu_\al\, t^\nu_\bt \nabla_\mu n^{\hi}_{\nu}$.

In any event, given eq.~\reef{gacod}, it is clear that the curvatures
in the two expressions for the horizon entropy agree when
$K^{\hi}_{\al\bt}=0$. Hence the two separate proposals will agree in
evaluating the horizon entropy for a stationary black hole with a
Killing horizon. Now a natural extension of eq.~\reef{define} to
Lovelock gravity would be that the holographic entanglement entropy
would be found by extremising the expression which yields black hole
entropy. Hence, in fact, eqs.~\reef{Waldformula2} and
\reef{Waldformula3} provide two natural candidates for the holographic
entanglement entropy. Further, as we will find below, in calculating
the holographic entanglement entropy, the relevant extrinsic curvatures
do not vanish in general and so these two expressions really provide
distinct proposals.

\subsection{Entanglement entropy and the trace anomaly} \label{EEcft}

We turn now to a CFT calculation of entanglement entropy, without
reference to holography. The results of these field theory calculations
will provide a benchmark against which we can compare our holographic
calculations of entanglement entropy. For a conformal field theory in
an even number of spacetime dimensions, the coefficient of the
universal term in the entanglement entropy can be determined through
the trace anomaly. This result relies on a common modification of the
usual replica trick \cite{cardy0} which is prevalent in the high energy
physics literature and which gives the calculations a geometric
character \cite{callan}. This `geometric approach' was first used to
establish the connection between entanglement entropy and the trace
anomaly for two-dimensional CFT's \cite{finn}. Later, similar results
were also found for higher dimensions in \cite{rt2,solo}. In the
following, we will not present the details of these calculations,
focusing on the results instead, and so we refer the interested reader
to \cite{cosmic} for a comprehensive discussion.

However, first let us recall the trace anomaly \cite{traca},
\be \langle\,T^i{}_i\,\rangle = \sum_n B_n\,I_n -2\,(-)^{d/2}A\, E_d +
B'\,\nabla_i J^i\,, \labell{trace} \ee
which defines the central charges for a CFT in an even number of
dimensions, $d=2p$. Each term on the right-hand side is a Weyl
invariant constructed from the background geometry. In particular,
$E_d$ is the Euler density in $d$ dimensions. Using the expressions
presented earlier in eq.~\reef{term0}, we write
$E_{2p}=\cL_{2p}(R)/[(4\pi)^p\, \Gamma(p+1)]$. This normalization
ensures that integrated over a $d$-dimensional sphere:
$\oint_{S^d}d^d\!x\sqrt{g}\, E_d =2$. A general construction of the
$I_n$ can be found in \cite{feffer}. In this approach, the natural
building blocks of these invariants are the Weyl tensor $C_{ijkl}$, the
Cotton tensor $C_{ijk}$ and the Bach tensor $B_{ij}$ (as well as
covariant derivatives of these). A useful observation is that these
basis tensors all vanish on a conformally flat background and hence,
\eg $I_n|_{S^d}=0$. Finally, the last term in eq.~\reef{trace} is a
conformally invariant but also scheme-dependent total derivative. That
is, this last contribution can be eliminated by the addition of a
finite and covariant counter-term to the effective action. In any
event, we note that these terms play no role in the following simply
because they are total derivatives. A final note on our
conventions\footnote{Our conventions are adopted from
\cite{cosmic,cthem} and so we refer the reader there for a more
detailed discussion.} is that the stress tensor is defined by
$T^{ij}\equiv -2/\sqrt{-g}\,\delta I/\delta g^{ij}$ in Minkowski
signature. However, in Euclidean signature, the sign is flipped to
$T^{ij}\equiv 2/\sqrt{g}\,\delta I_\mt{E}/\delta g^{ij}$.

Now consider calculating the entanglement entropy in the CFT using the
geometric approach mentioned above. First,\footnote{Actually the first
step in applying the replica trick is Wick rotate to Euclidean
signature.} a certain entangling surface $\Sigma$ is chosen which
divides the initial time slice into two separate regions, $V$ and
$\bar{V}$, as illustrated in figure \ref{pictx}. Following
\cite{rt2,solo}, we consider the variation of the entanglement entropy
under a uniform scale transformation of the system. This technique can
only be successfully applied when the geometry for which we are
calculating the entanglement entropy contains a single scale $\ell$.
Then the analysis of \cite{cosmic} leads to the following expression:
 \beq
\ell\,\frac{\partial S_\mt{EE}}{\partial \ell} = 2 \pi\, \int_{\Sigma}
d^{d-2}x\,\sqrt{h}\,\tilde{\veps}^{i j}\,\tilde{\veps}_{k l}\,
\left[\,\sum_n B_n\,\frac{\partial I_n}{\partial R^{i j}{}_{k l}}
-2\,(-)^{d/2}A\, \frac{\partial E_d}{\partial R^{i j}{}_{k l}}\
\right]\,.
 \labell{Waldformula9}
 \eeq
where $\tilde{\veps}_{ij}$ is the two-dimensional volume form in the
space transverse to $\Sigma$. Implicitly, for these computations, the
background geometry has Euclidean signature and hence
$\tilde{\veps}_{ij}\tilde{\veps}^{ij}=2$. The last term in this
expression can be further simplified using \cite{cosmic}
 \be
2 \pi\, \tilde{\veps}^{i j}\,\tilde{\veps}_{k l}\, \frac{\partial
E_d}{\partial R^{i j}{}_{k l}}=E_{d-2}(\,\R)\,.
 \labell{euler}
 \ee
That is, this contribution is replaced by the Euler density in
$d\!-\!2$ dimensions but constructed using the intrinsic curvatures on
$\Sigma$. This simplification relies on an implicit assumption in this
construction, which is that the extrinsic curvatures for $\Sigma$
vanish, and also uses eq.~\reef{gacod}. Now it is straightforward to
verify that the integral on the right-hand side of
eq.~\reef{Waldformula9} is scale invariant. Hence we can integrate with
respect to the scale $\ell$ to arrive at
 \beq
S_\mt{EE} = \log(\ell/\delta)\, \int_{\Sigma}
d^{d-2}x\sqrt{h}\,\left[\,2\pi\,\tilde{\veps}^{i j}\,\tilde{\veps}_{k
l}\ \sum_n B_n\,\frac{\partial I_n}{\partial R^{i j}{}_{k l}}
-2\,(-)^{d/2}A\, E_{d-2}\,\right]\,,
 \labell{Waldformula8}
 \eeq
where $\delta$ is the short-distance cut-off that we use to regulate
the calculations. Hence this calculation has identified the universal
contribution to the the entanglement entropy, \ie the term proportional
to $\log \delta$ in even $d$. Further, the above result shows that the
coefficient of this term is some linear combination of all of the
central charges, where the precise linear combination depends on the
geometry of the entangling surface $\Sigma$ and of the background
geometry \cite{rt2,solo}.
\FIGURE[t]{
\includegraphics[width=0.8\textwidth]{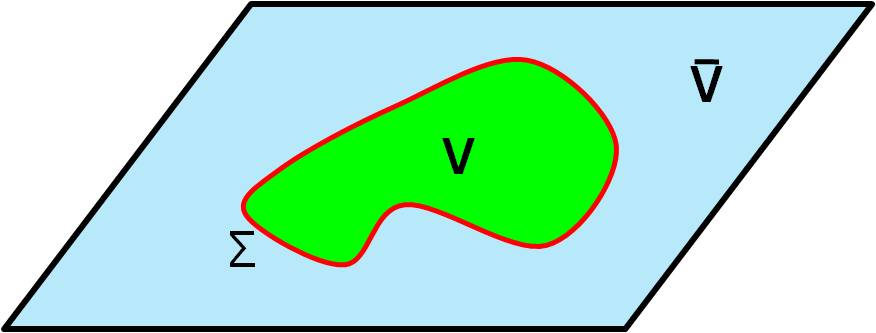}
\caption{Initial time slice divided into two regions $V$ and $\bar{V}$
by the entangling surface $\Sigma$.} \label{pictx}}

Let us add a few additional remarks about this calculation: As
commented above, an implict assumption in the above calculation is that
the extrinsic curvatures of the entangling surface vanish \cite{solo}.
Otherwise we should expect that additional `corrections' involving the
extrinsic curvature must appear in the final result. As we will see
below, these corrections were identified for $d=4$ in \cite{solo}.
However, there is, in fact, a stronger assumption at play in these
calculations. Namely, in applying the geometric approach to calculate
entanglement entropy, one should assume that there is a rotational
symmetry around the $\Sigma$ \cite{cosmic} and in fact, it is this
symmetry that ensures that $K^{\hi}_{ab}=0$. We will see in section
\ref{broken} that a new class of corrections (independent of the
extrinsic curvatures) can arise when the rotational symmetry is not
present. As a final note, we observe that if this calculation was
performed for a CFT in an odd number of spacetime dimensions, the
result would vanish because the trace anomaly is zero for odd $d$.
However, this is in keeping with the expectation that there is no
logarithmic contribution to the entanglement entropy for odd $d$ and
with a smooth entangling surface $\Sigma$.

We now explicitly apply eq.~\reef{Waldformula8} for $d=4$ and 6. These
results will then be the central consistency tests for our holographic
calculations of entanglement entropy for Lovelock gravity.

\subsubsection{Entanglement entropy for $d=4$} \label{eed4}

The trace anomaly for four dimensions is well studied and we present it
here using the more conventional nomenclature for the central charges:
 \be
\langle\, T^i{}_i \,\rangle=\frac{c}{16\pi^2}\,
C_{ijkl}C^{ijkl}-\frac{a}{16\pi^2}\left( R_{ijkl}R^{ijkl}-4R_{ij}R^{ij}
+ R^2\right) \,, \labell{trace4}
 \ee
where $C_{ijkl}C^{ijkl}= R_{ijkl}R^{ijkl}-2R_{ij}R^{ij}+R^2/3$ is the
square of the four-dimensional Weyl tensor and the second term is
proportional to the four-dimensional Euler density. We have discarded
the scheme-dependent total derivative in this expression, as it plays
no role in our analysis. Comparing to eq.~\reef{trace}, we have $a=A$
and $c=16\pi^2 B_1$ with $I_1=C_{ijkl}C^{ijkl}$. Now applying
eq.~\reef{Waldformula8}, the universal contribution to the entanglement
entropy becomes
 \beq
S_\mt{EE} = \log(\ell/\delta)\,\frac{1}{4\pi} \int_{\Sigma}
d^{2}x\sqrt{h}\,\left[\, c\, C^{i j k l}\,\tilde{\veps}_{i
j}\,\tilde{\veps}_{k l} -2\,a\, \R\,\right]\,,
 \labell{Waldformula8x}
 \eeq
where $\R$ is the Ricci scalar for the intrinsic geometry on $\Sigma$.

Similar expressions were previously derived in \cite{rt2,solo}. Our
results can be brought closer to the form presented there using the
formalism introduced in section \ref{wald}. We introduce two
orthonormal vectors $n^\hi_i$ (with $\hi=1,2$) which span the
transverse space to the entangling surface $\Sigma$. Then volume form
and metric in this space can be written as $\tilde\veps_{ij}=n^{\hat
1}_i n^{\hat 2}_j-n^{\hat 1}_j n^{\hat 2}_i$ and
$\tilde{g}^{\perp}_{ij} =\delta_{\hi\hj}\,n^\hi_i\,n^\hj_j$,
respectively. Now a useful identity which follows from these
definitions is
 \beq
\tilde\veps_{ij}\,\tilde\veps_{kl}=
\tilde{g}^{\perp}_{ik}\tilde{g}^{\perp}_{jl}
-\tilde{g}^{\perp}_{il}\tilde{g}^{\perp}_{jk}\,. \labell{useful6}
 \eeq
Using this identity, we can express our result for the universal
contribution to the entanglement entropy as
 \beq
S_\mt{EE} = \log(\ell/\delta)\,\frac{1}{2\pi} \int_{\Sigma}
d^{2}x\sqrt{h}\,\left[\, c\, C^{i j k l}\,\tilde{g}^\perp_{i
k}\,\tilde{g}^\perp_{j l} -\,a\, \R\,\right]\,,
 \labell{Waldformula8y}
 \eeq

As described above, this result can be reliably applied when there is a
rotational symmetry in the transverse space about $\Sigma$ (which
ensures that the extrinsic curvatures on $\Sigma$ vanish).
Ref.~\cite{solo} examined the possible corrections to
eq.~\reef{Waldformula8y} when $K^{\hi}_{ab}\not=0$. The extended result
which accounts for this possibility can be written \cite{solo}
 \beq
S_\mt{EE} = \log(\ell/\delta)\,\frac{1}{2\pi} \int_{\Sigma}
d^{2}x\sqrt{h}\,\left[\, c\, \left(C^{i j k l}\,\tilde{g}^\perp_{i
k}\,\tilde{g}^\perp_{j l} - K^\hi_a{}^b K^\hi_b{}^a+\frac12 K^\hi_a{}^a
K^\hi_b{}^{\,b}\right) -\,a\, \R\,\right]\,.
 \labell{Waldformula8yy}
 \eeq
It is interesting to note that a holographic calculation was used in
\cite{solo} to fix the final coefficients of the extrinsic curvature
terms in this expression. For comparison purposes in section \ref{EE4},
it is also useful to write this expression as
 \beq
S_\mt{EE} = \log(\ell/\delta)\,\frac{1}{2\pi} \int_{\Sigma}
d^{2}x\sqrt{h}\,\left[\, c\, \left(C^{abcd}\,h_{a c} \,h_{bd} -
K^\hi_a{}^b K^\hi_b{}^a+\frac12 K^\hi_a{}^a K^\hi_b{}^{\,b}\right)
-\,a\, \R\,\right]\,,
 \labell{Waldformula8yz}
 \eeq
where $h_{ab}$ is the induced metric on $\Sigma$. The equivalence of
the two expressions in eqs.~\reef{Waldformula8yy} and
\reef{Waldformula8yz} follows because the Weyl tensor is traceless, \ie
$C^{i j k l}\,g_{ik}=0$ and we can express the induced metric as a bulk
tensor with $h_{ij}=g_{ij}-\tilde{g}^\perp_{i k}$.

Further in \cite{solo}, this result was applied to evaluate the
entanglement entropy for various surfaces embedded in flat space. In
this case, the Weyl curvature vanishes and the entanglement entropy is
determined entirely by the contributions coming from the extrinsic
curvatures and from the intrinsic Ricci scalar. Considering the case
where the entangling surface is a two-sphere of radius $R$, one finds
that the two extrinsic curvature terms cancel. The entanglement entropy
\reef{Waldformula8yz} then becomes
 \beq
S_\mt{EE} = -4\,a\,\log(R/\delta)\,,
 \labell{Wald2sphere}
 \eeq
where we have substituted $R$ as the relevant scale $\ell$. Another
simple case to consider is when the entangling surface is chosen to be
an infinite cylinder, in which case the intrinsic curvature vanishes.
If we let the radius of the cylinder be $R$ and we introduce regulator
scale $H$ along the length of the cylinder, the entanglement entropy
\reef{Waldformula8yz} then becomes
 \beq
S_\mt{EE} = -\frac{c}2\,\frac{H}{R}\,\log(R/\delta)\,.
 \labell{Wald2cyl}
 \eeq
Hence with these two choices for the entangling surface, we are able to
isolate the two central charges with a calculation of the entanglement
entropy.

In section \ref{EE4}, we will use the above results to test our
proposal for holographic entanglement entropy in Lovelock gravity. One
comment, perhaps worth making at this point, is that while the
derivation of eq.~\reef{Waldformula8yz} did account for the possibility
that the extrinsic curvature was nonvanishing \cite{solo}, no
consideration was given to whether the transverse space to $\Sigma$
possessed a rotational symmetry. As we discuss in section
\ref{discuss}, the latter does not seem to lead to any difficulties in
$d=4$.

\subsubsection{Entanglement entropy for $d=6$}

In six dimensions, the trace anomaly \reef{trace} can be explicitly
written as \cite{tseytlin}
 \be
\langle\,T^i{}_i\,\rangle=\sum_{n=1}^3 B_n\, I_n + 2 A \,E_6
\labell{trace6}
 \ee
where
 \bea
I_1&=&C_{k i j l}C^{i m n j}C_{m~~\,n}^{~~k l} ~, \qquad
I_2=C_{i j}^{~~k l}C_{k l}^{~~m n}C^{~~~i j}_{m n}~,\nonumber \\
I_3&=&C_{i k l m}(\nabla^2 \, \de^{i}_{j}+4R^{i}{}_{j}-{6 \over 5}\,R
\,
\de^{i}_{j})C^{j k l m}\,,\labell{trace6x}\\
E_6&=&\frac{1}{384\pi^3}\cL_6 \nonumber
 \eea
with $\cL_6$ defined in eq.~\reef{term0}. We also explicitly write out
$\cL_6$ in eq.~\reef{euler6}. In eq.~\reef{trace6}, we have again
discarded the scheme-dependent total derivative. The above choice for
the basis of the conformal invariants has the virtue that the
non-topological terms $I_n$ all vanish when evaluated on a conformally
flat space. Now from eq.~\reef{Waldformula8}, the universal
contribution to the entanglement entropy becomes
 \beq
S_\mt{EE} = \log(\ell/\delta)\, \int d^{4}x\sqrt{h}\,\left[\,2\pi\,
\sum_{n=1}^3 B_n\,\frac{\partial I_n}{\partial R^{i j}{}_{k
l}}\,\tilde{\veps}^{i j}\,\tilde{\veps}_{k l} +2\,A\,
E_{4}\,\right]_{\Sigma}\,,
 \labell{Waldformula8z}
 \eeq
where
 \bea
  \frac{\partial I_1}{\partial R^{i j}{}_{k l}}\,\tilde{\veps}^{i
j}\,\tilde{\veps}_{k l} &=& 3 \left(C^{j m n k}\, C_{m~~n}^{~~i
l}\,\tilde\veps_{i j}\,\tilde\veps_{k l} - \frac{1}{4}\,
C^{iklm}\,C^{j}_{~kl m }\,\tilde{g}^{\perp}_{i j} + \frac{1}{20}\,
C^{ijkl}\,C_{ijkl}\right)
 \,,\nonumber \\
\frac{\partial I_2}{\partial R^{i j}{}_{k l}}\,\tilde{\veps}^{i
j}\,\tilde{\veps}_{k l} &=&3\left(C^{k l m n}\, C_{m n}^{~~~ i j}\,
\tilde\veps_{i j}\,\tilde\veps_{k l} -C^{iklm}\,C^{j}_{~kl m
}\,\tilde{g}^{\perp}_{i j} + \frac{1}{5}\, C^{ijkl}\,C_{ijkl}\right)
 \,,\labell{items9} \\
\frac{\partial I_3}{\partial R^{i j}{}_{k l}}\,\tilde{\veps}^{i
j}\,\tilde{\veps}_{k l} &=&  2\left(\Box\, C^{i j k l}+ 4\,R^{i}{}_{m}
C^{m j k l}-\frac{6}{5}\, R\, C^{i j k l}\right) \tilde\veps_{i
j}\,\tilde\veps_{k l}- 4\, C^{i j k l}\,R_{ik}\,\tilde{g}^{\perp}_{jl}
 \nonumber \\
&& \qquad+ 4\,C^{iklm}\,C^{j}_{~kl m }\,\tilde{g}^{\perp}_{i
j}-\frac{12}{5}\, C^{ijkl}\,C_{ijkl} \, . \nonumber
 \eea
The expressions above have been simplified using the identities:
$\tilde{\veps}_{ij}\,\tilde{\veps}^{ij}=2$ and $\tilde{g}^{\perp}_{ik}=
\tilde{\veps}_{i j}\,\tilde{\veps}_{k l}\,g^{jl}$.

As described above, this result can be reliably applied for entangling
surfaces with rotational symmetry in the transverse space, in which
case there is zero extrinsic curvature. In section \ref{EE6}, we will
explicitly evaluate eq.~\reef{Waldformula8z} for various surfaces
satisfying these constraints to test our proposal for holographic
entanglement entropy in Lovelock gravity. It would, of course, be
interesting to extend the above expression \reef{Waldformula8z} to
account for the possibility that the extrinsic curvature is
nonvanishing, following the approach of \cite{solo} in $d=4$. However,
one quickly realizes extending these calculations to $d=6$ is arduous
task and further, we will show in section \ref{EE6} that there are
other corrections unrelated to the extrinsic curvature.

\section{Not Wald entropy!} \label{not}

Our goal is to understand how to compute the holographic EE in the
presence of higher curvature interactions in the bulk theory. A
discussed in the introduction it seems that as long as the prescription
is one of minimizing a `surface functional', we can expect that the
holographic EE will satisfy subadditivity, as in \cite{head2}. Further,
the close connection of entanglement entropy and black hole entropy
suggests that the new functional must coincide with the expression for
Wald entropy \reef{Waldformula} when evaluated on a black hole horizon.
The simple suggestion would then be that we extend eq.~\reef{define} to
higher curvature gravity by extremizing precisely the Wald formula over
the bulk surfaces homologous to the boundary region of interest. The
recent discussion of \cite{circle4} on spherical entangling surfaces
would seem to lend some credence to this prescription.

Unfortunately, we can easily show that the na\"ive first guess above
for the extension of eq.~\reef{define} to higher curvature theories
simply fails to provide the correct holographic EE for general
entangling surfaces. For this purpose, we will use two results that
were originally derived in \cite{cosmic,cthem}. We only present these
results here and refer the interested reader to \cite{cosmic} for
further details. We can keep the discussion general and do not need to
specify the gravitational theory, beyond that it has a covariant
action. Then, in an AdS$_{d+1}$ background, the gravitational equations
of motion yield
  \be
\left.\frac{\delta \cal L}{\delta
R^{\mu\nu}{}_{\rho\sigma}}\right|_{AdS}= -{\tilde
L^2\over4\,d}\left.{\cal L}\right|_{AdS}\,\left(\delta_\mu{}^\rho\,
\delta_\nu{}^\sigma -\delta_\mu{}^\sigma\, \delta_\nu{}^\rho\right)\,.
 \labell{adseom2}
 \ee
Further, motivated by the short-cut to calculating the $A$-type trace
anomaly \cite{adam}, the authors of \cite{cosmic,cthem} found
 \be
\ads\equiv-\frac{\pi^{d/2}\,\tilde L^{d+1}}{d\,\Gamma\left(d/2 \right)}
\left.{\cal L}\right|_{AdS}\,.
 \labell{adamA2}
 \ee
Here $\ads$ is a specific central charge characterizing the
$d$-dimensional boundary CFT.\footnote{It was also shown that $\ads$
provides a measure of the density of the degrees of freedom in the
boundary CFT \cite{cosmic,miguel}.} The above expression generalizes
eq.~\reef{adamA} for odd or even $d$. Recall that for even $d$, we have
$\ads=A$, the coefficient of the $A$-type trace anomaly. Now our
candidate for the holographic EE is to extremize the Wald entropy
\reef{Waldformula} evaluated on ($d$--1)-dimensional surfaces $m$
homologous to the boundary region of interest. For simplicity, we will
test this proposal in pure AdS$_{d+1}$ space, which will suffice to
consider the cases where the entangling surface is embedded in flat
Minkowski space $R^{1,d-1}$, a cylindrical background $R\times S^{d-1}$
or any number of conformally flat backgrounds --- see, \eg
\cite{faces}. Hence we consider evaluating the expression for Wald
entropy on some bulk surface $m$,
 \beq
S_\mt{W} = -2 \pi \int_m d^{d-1}x\sqrt{h}\
\frac{\partial{\cL}}{\partial R^{\mu\nu}{}_{\rho\sigma}}\,
\hat{\veps}^{\,\mu\nu}\,\hat{\veps}_{\rho\sigma}=\frac{2\pi}{\pi^{d/2}}
\,\frac{\Gamma(d/2)}{\tilde L^{d-1}} \,\ads\ \int_m d^{d-1}x\sqrt{h}
\,.
 \labell{tryWald}
 \eeq
We used eqs.~\reef{adseom2} and \reef{adamA2} to produce the second
expression on the right above. Hence with this prescription, the
calculation of holographic EE would again reduce to extremizing the
area of the bulk surface. Further the entire result would always be
proportional to the central charge $\ads$, independent of the choice of
the entangling surface $\Sigma$. In particular, for even $d$, the
coefficient of the logarithmic contribution would be proportional to
$A$. However, this result is simply incorrect. As we saw in section
\ref{EEcft}, field theoretic calculations indicate that this universal
contribution to the entanglement entropy is proportional to a linear
combination of all of the central charges appearing in the trace
anomaly \reef{trace}. Further the specific linear combination appearing
here depends on the geometry of the entangling surface $\Sigma$ and of
the background in which $\Sigma$ is embedded.

Hence the proposal that holographic EE would be calculated by
extremizing the Wald entropy clearly contradicts the general
expectations from purely CFT calculations. In the next two sections, we
focus our discussion on Lovelock gravity. While Wald entropy is ruled
out, in this case, there remains a second natural candidate in
eq.~\reef{Waldformula3} for the new functional with which to calculate
holographic EE. In the following, we will verify the proposal that
extremizing the expression for $S_\mt{JM}$ over the bulk surfaces $m$
will properly determine the EE for the holographic CFT's dual to
Lovelock gravity.

\section{EE for $d=4$ holographic CFT} \label{EE4}

Here we focus on the case of a four-dimensional boundary theory. In
this case with Lovelock gravity in a five-dimensional bulk, only the
curvature-squared interaction contributes to the action \reef{GBactg}
leaving
 \beq
I = \frac{1}{2\lp^3} \int \mathrm{d}^5x \, \sqrt{-g}\, \left[
\frac{12}{L^2} + R + \frac{\lambda L^2}{2}\cL_4 \right]\,.
 \labell{GBact2}
 \eeq
Comparing to the notation of section \ref{glove}, we have
$\lambda=\lambda_2=2c_2$ and explicitly evaluating $\cL_4$ using
eq.~\reef{term0} yields
 \beq
\cL_4=R_{\mu\nu\rho\sigma}R^{\mu\nu\rho\sigma}-4\,
R_{\mu\nu}R^{\rho\sigma}+R^2\,. \labell{euler4}
 \eeq
In this case, eq.~\reef{roots} reduces to a simple quadratic equation
for which the physical (ghost-free) root is
\be \fin=\frac{1-\sqrt{1-4\lambda}}{2\lambda}\,. \labell{root1} \ee
The two central charges appearing in the trace anomaly \reef{trace4}
can be calculated using the techniques of \cite{sken} yielding
\cite{old2}
 \be
c= \pi^2\frac{\tL^3}{\lp^3}\left(1-2\la\fin \right)\ ,
 \qquad
a=\pi^2\frac{\tL^3}{\lp^3}\left(1-6\la\fin  \right)\ .
 \labell{ccaa}
 \ee

Now we would like to test the proposal that the holographic
entanglement entropy in this theory is determined by extremizing the
expression in eq.~\reef{Waldformula3} over the bulk surfaces $m$
homologous to the appropriate boundary region. Explicitly evaluating
this functional for the present case yields
 \beq
S_\mt{JM} = \frac{2 \pi}{\lp^3} \int_m d^3x\sqrt{h}\,\left[ 1+
\lambda\,L^{2}\,\R \right]+\frac{4\pi}{\lp^{3}} \int_{\partial m}
d^{2}x\sqrt{h}\,\lambda\,L^{2}\,\K \,,
 \labell{Waldformula3x}
 \eeq
where $\R$ denotes the Ricci scalar for the intrinsic geometry on $m$.
Similarly, $\K$ denotes the trace of the extrinsic curvature of the
boundary $\partial m$. We have added this `Gibbons-Hawking' boundary
term in eq.~\reef{Waldformula3x} to provide a good variational
principle in extremizing this functional. In the rest of this section,
we test our hypothesis by evaluating the holographic entanglement
entropy and comparing to the general results derived of any CFT, which
are presented in section \ref{eed4}. We will first consider the cases
where the entangling surface is a sphere\footnote{The case of the
sphere can be analyzed without restricting $d$ to a particular value.
This analysis is presented in Appendix \ref{appx:EE-disk} } and an
infinite cylinder and our holographic results will match precisely with
the CFT results in eqs.~\reef{Wald2sphere} and \reef{Wald2cyl}. We
conclude this section by showing that by applying the techniques
developed in \cite{adam} to the functional (\ref{Waldformula3x}), in
fact, we can recover the general result \reef{Waldformula8yz} for the
EE with any (smooth) entangling surface.

\subsection{EE of the sphere}

In this case, it is convenient to parameterize the $AdS_5$ metric as
follows
 \be
 ds^2=\frac{\tilde L^2}{z^{2}}\left(dz^2-dt^2+dr^2+r^2d\Omega^2_{2}
 \right)
 \,.\labell{metric99x}
 \ee
Recall that the AdS scale is given by $\tilde L^2=L^2/\fin$. Here, the
asymptotic boundary is approached with $z\to0$ and as usual, we
regulate the evaluation of eq.~\reef{Waldformula3x} by introducing a
short distance regulator with $z=z_{min}=\delta$. Of course, with this
choice of coordinates, the boundary metric is simply flat Minkowski
space in spherical polar coordinates. We will calculate the
entanglement entropy for the interior of a sphere $r=R$ on the $t=0$
surface in the AdS boundary.

As shown in Appendix \ref{appx:EE-disk}, the surface that minimizes
(\ref{Waldformula3x}) can be parameterized as
 \be
 r(\theta)=R\,\cos\theta~ , \quad z(\theta)=R\,\sin\theta~, \quad
 \delta/R\leq \theta \leq {\pi \over
 2}~. \labell{range99x}
 \ee
Upon evaluating eq.~\reef{Waldformula3x} for this surface, the leading
term is a non-universal contribution proportional to $R^2/\delta^2$.
However, if we focus on the universal logarithmic contribution, we find
 \be
S_{\mt{JM}}=-  4\pi^{2} \frac {\tilde
L^3}{\lp^3}\[1-6\,\lambda\,\fin\]\,\log (R/\delta)+\ldots
  \labell{EEdisk99x}
 \ee
Moreover, given the central charges in eq.~(\ref{ccaa}), we see that
this result is proportional to $a$ and the result \reef{EEdisk99x} can
be expressed as
 \be
 S_{\mt{JM}}=-4\,a\,\log(R/\delta) + \cdots \,.
 \ee
With the ellipsis, we are denoting the power-law divergent and finite
terms. This result agrees precisely with that given in
eq.\reef{Wald2sphere}, which was derived from purely CFT techniques. At
this point, let us also observe that in our calculation of the
entanglement entropy, the surface term in eq.~\reef{Waldformula3x} does
not contribute to the above logarithmic term \reef{EEdisk99x}.

\subsection{EE of the cylinder.}
\label{sec:EEcyl}

In the case of the cylinder, we choose the following coordinates to
parameterize the $AdS_5$ space
 \be
 ds^2=\frac{\tilde L^2}{z^{2}} \left(dz^2-dt^2+dx^2+dr^2+r^2d\phi^2
 \right)\, .
 \ee
With this choice of coordinates, the boundary metric is again flat
Minkowski space now in cylindrical polar coordinates. We choose the
entangling surface as the cylinder $r=R$ on the $t=0$ surface in this
boundary geometry. In the following, we also introduce a regulator
length $H$ for the $x$ direction, \ie along the length of the cylinder.
The rest of our notation is the same as in the previous subsection.

In evaluating the functional \reef{Waldformula3x}, let us parameterize
the surface $m$ with $r(z)$. We have made some general analysis for a
cylindrical entangling surface in a $d$-dimensional boundary theory in
Appendix \ref{cylinder99x}. Reducing these results to $d=4$, we arrive
at
 \be
S_{\mt{JM}}=\frac {4\pi^2\,\tilde L^3\,H}{\lp^{\,3}}
\int_\delta^{z_{max}} dz \,{r \over z^3}\,\sqrt{r'^2+1}\,\left(1
+2\,\fin \,\lambda\,{1-zr'/r \over 1+r'^2}
 \right)~,
 \label{eqn:EE-4dcyldr}
 \ee
where $z_{max}$ denotes the maximal radius which the surface $m$
reaches. Extremizing the above expression leads to the following
`equation of motion'
 \bea
&&z(1-2\fin\lambda)+r'[(1+4\fin\lambda)z
r'+3r(1-2\fin\lambda+r'^2)]
 \labell{eom99x}\\
&&\qquad\qquad={r''\,z\over 1+r'^2}\left(6\fin\lambda
zr'+r(1-2\fin\lambda+(1+4\fin\lambda)r'^2)\right)
 ~.\nonumber
 \eea
To identify the universal contribution in eq.~\reef{eqn:EE-4dcyldr}, it
suffices to solve this equation asymptotically by substituting the
expansion
 \be
 r(z)=r_0+r_1\, z + r_2\, z^2 + \cdots\,, \labell{expand99x}
 \ee
which yields
 \bea
 r_1&=&0~,
 \non
 r_2&=&-{1 \over 4 r_0}
 ~.
 \eea
Applying the boundary condition $r(z=0)=R$, we find
 \be
 r(z)=R\,\left(1- {z^2\over 4 R^2}+ \cdots\right)
 \ .
 \ee
Substituting this asymptotic expansion back into the expression for the
EE (\ref{eqn:EE-4dcyldr}) and using the results for the central charges
(\ref{ccaa}), we finally obtain
 \be
 S_{\mt{JM}}=-\frac{c}2\,{H\over R}\log(R/\delta)+\cdots\,,
 \labell{cylresult99x}
 \ee
where ellipsis again denotes the finite and nonuniversal contributions.
Once again, our computation of the holographic EE using
eq.~\reef{Waldformula3x} is in precise agreement with general result
\reef{Wald2cyl} for the universal logarithmic contribution.

\subsection{General case: EE as the Graham-Witten anomaly} \label{sch4d}

Here, we consider general (smooth) entangling surfaces $\Sigma$ in the
boundary CFT and apply the methods developed in \cite{adam} to
evaluating eq.~(\ref{Waldformula3x}). This general formalism allows the
holographic evaluation of trace anomalies for submanifolds, \ie
Graham-Witten anomalies \cite{grwit}. Further the approach rests on the
so-called Penrose-Brown-Henneaux (PBH) transformations, which
correspond to the subgroup of bulk diffeomorphisms which generate Weyl
transformations of the boundary metric. This approach enables us to
perturbatively evaluate the metric and the shape of the minimal surface
in the vicinity of the AdS boundary without resorting to either the
gravitational equations of motion or to the short distance cutoff
$\delta$. In fact, this feature completely fixes the necessary geometry
for $d=4$, whereas for higher dimensions, one still needs to consider
the equations of motion in order to fix various constants which cannot
be determined on the basis of PBH transformations.

We start from a brief review of the general method and the interested
reader can find the details in the original papers \cite{adam}. We
denote the dimensions of the AdS boundary and of the submanifold
$\Sigma$ embedded in the boundary as $d$ and $k$, respectively. For the
initial discussion, we leave $d$ and $k$ as general and however, at the
end of the discussion, we will focus on $d=4$ and $k=2$, as is relevant
for the holographic EE here.

In the Fefferman-Graham (FG) gauge, coordinates are chosen for the bulk
metric \cite{feffer}
 \be
ds^2=G_{\mu\nu}dX^\mu dX^\nu={\tilde L^2 \over 4} \left({d\rho\over
\rho}\right)^2 +{1\over\rho}\,g_{ij}(x,\rho)dx^i dx^j\,.
 \labell{FGgauge}
 \ee
where $g_{i j}(x,\rho)$ admits a Taylor series expansion in the radial
coordinate $\rho$:
 \be
 g_{ij}(x,\rho)=\gz_{ij}(x)+\go_{ij}(x)\,\rho
 +\gs_{ij}(x)\,\rho^2+\cdots\,.
 \labell{expandg3}
 \ee
The asymptotic AdS boundary is approached with $\rho\to0$. The first
term in this expansion, $g_{ij}(x,0) =
\overset{\scriptscriptstyle{(0)}} g_{ij}(x)$ is identified with the
background metric of the dual CFT. Exploring the transformation
properties of the $g_{ij}(x,\rho)$ under the PBH diffeomorphisms, which
preserve the FG gauge, one can essentially determine the remaining
coefficients in the Taylor series for $n<d/2$ --- see appendix
\ref{schwimmer} for further details. The embedding of the
($k+1$)-dimensional submanifold $m$ into the $(d+1)$-dimensional bulk
is described by $X^\mu=X^\mu(y^{a},\tau)$, where $X^\mu=\lbrace
x^i,\rho\rbrace$ are the bulk coordinates and $\sigma^\al= \lbrace
y^a,\tau\rbrace$ are the coordinates on surface $m$ (with $a=1,..,k$).
Reparameterizations of $m$ can be fixed by imposing
 \be
 \tau=\rho\qquad{\rm and}\qquad h_{a\tau}=0
 \, ,
 \labell{eqn:gauge-brane}
 \ee
where $h_{\al\bt}$ denotes the induced metric on $m$.

By definition, the PBH transformations preserve the FG gauge
\reef{FGgauge}, however, they do change $\rho$ in general. Thus to stay
within the above gauge \reef{eqn:gauge-brane}, one needs to apply the
compensating world-volume diffeomorphism on $m$. The requirement of
preserving eq.~\reef{eqn:gauge-brane}uniquely fixes the induced (by the
PBH transformation) world-volume diffeomorphism and the transformation
rule of the embedding functions $X^\mu(y^a,\tau)$. Let us make a Taylor
expansion or the embedding functions in $\tau$,
 \be
 X^i(\tau,y^a)=\Xz^i(y^a)+ \Xo^i(y^a)\,\tau+\cdots\,,
 \label{braneform}
 \ee
where $\overset{\scriptscriptstyle{(0)}}{X^i}(y^a)$ describes the
position of $\del m$ on the boundary of AdS. In the case of interest,
this matches the position of the entangling surface $\Sigma$ in the
boundary metric $\overset{\scriptscriptstyle{(0)}} g_{ij}(x)$. Now
studying the above transformation rules order by order, one can
determine the higher coefficients, \eg
 \be
  \Xo^i(y^a)={\tilde L^2\over 2k}K^i(y^a)\,, \labell{nextX}
 \ee
with $K^i$ being the trace of the second fundamental form of the
boundary submanifold $\Sigma$, \ie $K^i=n^{i }_{\hj}\,K^{\hj}_{a
b}\overset{\scriptscriptstyle{(0)}}{h}{}^{ab}$ where
$\overset{\scriptscriptstyle{(0)}}{h}_{ab}$ is the induced metric on
$\Sigma$ --- see below.\footnote{Note that we are adopting the notation
of \cite{adam} here by contracting the extrinsic curvatures with a
normal vector, \ie $K^i_{ab}=n^{i }_{\hj}\,K^{\hj}_{ab}$. Hence in the
following formulae, the extrinsic curvatures carry a coordinate index
$i$, rather than a frame index $\hi$, as in our previous expressions.}
As a result, the induced metric on $m$
%and the world-volume diffeomorphisms $(\tilde\xi^\tau, \tilde\xi^a)$
compatible with the gauge choice (\ref{eqn:gauge-brane}) is also
determined as
 \be
 h_{\tau\tau}={\tilde L^2\over4\tau^2}\Big(1+{\tilde L^2\over
 k^2}\,
 K^iK^j\gz_{ij}\,\tau+\cdots\Big)\,,\qquad
 h_{ab}={1\over \tau}\(\hz_{ab}+\ho_{ab}\,\tau+\cdots\)\, ,
\labell{inducemet}
 \ee
with
 \be
\hz_{ab}=\del_a\!\Xz^i\,\del_b\!\Xz^j\gz_{ij}\qquad{\rm and} \qquad
 \ho_{ab}=\go_{ab} -{\tilde L^2\over k} K^iK^j_{ab}\gz_{ij}
 \,.
 \labell{eqn:4d-ind-tautau}
 \ee
An explicit expression for $\overset{\scriptscriptstyle{(1)}}{g}_{ab}$,
which appears in the last formula, can be found in appendix
\ref{schwimmer}.

At this point, we set the dimensions to $d=4$  and $k=2$. Now applying
the above results, we find the following expansion for the intrinsic
Ricci scalar on $m$,
 \be
 \R=-{6\over \tilde L^2}+\left(\R_\Sigma+{2 \over \tilde L^2}
 \hz^{ab}\go_{ab}-{1\over2} K^i K^j\gz_{ij}\right)\tau+\cdots
 \labell{useful8}
 \ee
where $\R_\Sigma$ is the intrinsic curvature scalar on the boundary
surface $\Sigma$, \ie the entangling surface. Next using a variant of
Gauss-Codazzi relation \cite{mtw}, we re-express the term in
parentheses above as
 \be
\R_\Sigma+{2 \over \tilde L^2} \hz^{ab}\go_{ab}-{1\over2} K^i
K^j\gz_{ij} = \hz^{ac}\hz^{bd} C_{abcd}-(\mathrm{tr}(K^i
K^j)-{1\over2}K^i K^j)\gz_{ij} \,.
 \labell{useful9}
 \ee
Now we combine all of these results together in evaluating $S_\mt{JM}$,
our surface functional for the holographic EE in
eq.~\reef{Waldformula3x}. Note that the asymptotic expansions above
suffice in identifying the logarithmic contribution and we find using
eq.~(\ref{ccaa}),
 \be
S_{EE}={\log(\ell/ \delta) \over 2\pi} \int_{\Sigma}d^2x
(\hz)^{1/2}\left[\, c\,\Big(\hz^{ac}\hz^{bd}C_{abcd}-\mathrm{tr}(K^i
K^j)+{1\over2}K^i K^j\Big)-a \, \R_\Sigma\right]
  + \cdots\,, \labell{bigresult}
 \ee
where $\ell$ is some macroscopic scale that emerges from the CFT
geometry. Now if we account for the slightly different notation here
and in section \ref{eed4}, we see that this holographic result for an
arbitrary (smooth) entangling surface $\Sigma$ precisely matches the
universal entropy term \reef{Waldformula8yz} derived from purely field
theoretic considerations. Hence this final test seems a strong
indication that $S_\mt{JM}$ is the correct surface functional to
replace the area in eq.~\reef{define} when defining holographic EE in
Lovelock gravity.

As a final comment, we note that the boundary term which we added to
eq.~\reef{Waldformula3x} only contributes to power-law divergent and
finite terms in the holographic EE and does not contribute to the
universal term \reef{bigresult}.

\section{EE for $d=6$ holographic CFT} \label{EE6}

We now turn to the case of a six-dimensional boundary theory. In this
case with a seven-dimensional bulk, the curvature-squared and -cubed
interactions contribute to the Lovelock action \reef{GBactg} yielding
 \beq
I = \frac{1}{2\lp^5} \int \mathrm{d}^7x \, \sqrt{-g}\, \left[
\frac{30}{L^2} + R + \frac{L^2}{12}\,\lambda\, \cL_4(R) - \frac{
L^4}{24}\, \mu\,\cL_6(R)\right]\ .
 \labell{GBact3}
 \eeq
Comparing to the notation of section \ref{glove}, we have
$\lambda=\lambda_2=12c_2$ and $\mu=\lambda_3=-24c_3$. Further, $\cL_4$
is given in eq.~\reef{euler4} while explicitly evaluating $\cL_6$ using
eq.~\reef{term0} yields
 \beqa
\cL_6 &=& 4\, R_{\mu\nu}^{\,\,\,\,\,\,\rho\sigma}
R_{\rho\sigma}^{\,\,\,\,\,\,\tau\chi} R_{\tau\chi}^{\,\,\,\,\,\,
\,\,\mu\nu}-8\, R_{\mu\,\,\,\nu}^{\,\,\,\rho\,\,\,\,\sigma}
R_{\rho\,\,\,\,\sigma}^{\,\,\,\tau\,\,\,\chi}
R_\tau{}^\mu{}_\chi{}^{\nu} -24\, R_{\mu\nu \rho\sigma} R^{\mu\nu
\rho}{}_{\tau} R^{\sigma\tau} +3\, R_{\mu\nu\rho\sigma}
R^{\mu\nu\rho\sigma} R
 \nonumber\\
&&\qquad\quad
 +24\,R_{\mu\nu\rho\sigma} R^{\mu\rho}R^{\nu\sigma}+16\, R_\mu^{\,\,\nu}
R_\nu^{\,\,\rho} R_\rho^{\,\,\mu} -12\, R_\mu^{\,\,\nu} R_\nu^{\,\,\mu}
R + R^3\,, \labell{euler6}
 \eeqa
Substituting $d=6$ into eq.~\reef{roots} yields the cubic equation
 \be
1= \fin-\fin^2 \lambda -\fin^3 \mu\,.
 \ee
In principle, we can again solve for $\fin$ analytically, however, the
precise expression will not be needed in the following. Note that
implicitly we choose the particular root (the smallest positive root)
which gives the physical vacuum, as discussed in detail in \cite{old1}.
With $d=6$, there are four central charges appearing in the trace
anomaly, as discussed in section \ref{EEcft}. The holographic
expressions for the central charges were calculated in \cite{jan3}:
 \beqa
B_1&=&\frac{\tilde{L}^5}{\lp^5}\frac{-9 + 26 \fin \lambda + 51\fin^2 \mu}{288}\,,
\nonumber\\
B_2&=&\frac{\tilde{L}^5}{\lp^5}\frac{-9+34 \fin\lambda +75 \fin^2 \mu}{1152}\,,
\labell{center6} \\
B_3&=& \frac{\tilde{L}^5}{\lp^5}\frac{1-2\fin \lambda - 3\fin^2\mu}{384}\,,
\nonumber\\
A  &=& \pi^3\frac{\tilde{L}^5}{\lp^5}\frac{3 - 10 \fin \lambda - 45
\fin^2 \mu}{6} \, . \nonumber
 \eeqa
Of course, this expression for $A$ agrees with the general expression
given in eq.~\reef{llA}.

Now we would like to further test the proposal that the holographic EE
in Lovelock gravity is given by extremizing the expression in
eq.~\reef{Waldformula3} over the bulk surfaces $m$ homologous to the
appropriate boundary region. For seven-dimensional Lovelock gravity,
eq.~\reef{Waldformula3} becomes
 \beq
S_\mt{JM} = \frac{2 \pi}{\lp^{5}} \int_m d^{5}x\sqrt{h}\,\left[ 1+
\frac{\lambda}{6}\,L^{2}\,\R -
\frac{\mu}{8}\,L^4\,\left(\R_{\mu\nu\rho\sigma}
\R^{\mu\nu\rho\sigma}-4\R_{\mu\nu}\R^{\mu\nu} +\R^2\right)\right]\,.
 \labell{Waldformula3z}
 \eeq
While we could add an appropriate set of surface terms \cite{surf}, we
will not consider their contributions in the following. The focus of
our analysis will be the universal logarithmic contributions but, as
found in the previous section with $d=4$, the surface terms will only
make power-law contributions in the short distance cut-off $\delta$.
That is, they only contribute power-law divergent or finite terms in
the entanglement entropy.

Following the strategy of the previous section \ref{EE4}, we extract
the universal log-term in the holographic calculation and compare with
the corresponding CFT result \reef{Waldformula8z}. As previously noted,
the latter result can only be reliably applied with the entangling
surfaces where there is a rotational symmetry in the transverse space
\cite{cosmic}. Hence to test the proposal for holographic EE, we start
by applying it to various surfaces which possess the desired rotational
symmetry. We will find that in these cases there is full agreement
between \reef{Waldformula3z} and the general CFT results
\reef{Waldformula8z}. However, we note that the restriction for the CFT
analysis in section \ref{EEcft} is commonly stated as demanding that
the extrinsic curvature of the entangling surface should vanish, \eg
\cite{solo}. Hence we also examine the holographic EE for surfaces with
zero extrinsic curvature but without a rotational symmetry in the
transverse space. In this case we find a discrepancy between
eqs.~\reef{Waldformula3z} and \reef{Waldformula8z}. We argue that
eq.~\reef{Waldformula8z} is incomplete, \ie unable to properly
determine the universal contribution, for these cases. However, we are
able to use holography to construct the additional curvature terms
which must be added to eq.~\reef{Waldformula8z} to correctly determine
the universal EE.

\subsection{Entangling surfaces with rotational symmetry}
\label{unbroken}

As noted above, the CFT results of section \ref{EEcft} are only
reliable when the entangling surface has a rotational symmetry in the
transverse space \cite{cosmic}.  This symmetry is not generally present
when time is singled out in geometries of the form $R_t \times
\M_{d-1}$. Rather the Euclidean\footnote{Recall that these CFT
calculations are performed after Wick rotating to Euclidean signature.}
background must have a high degree of symmetry. As simple example, we
could take the six-dimensional boundary geometry to be $S^6$ and the
desired rotational symmetry would result by choosing $\Sigma$ as a
maximal $S^4$ within this geometry. A natural question would be: what
is the interpretation of the resulting entanglement entropy? A simple
Wick rotation of the $S^6$ back to Minkowski signature would yield
six-dimensional de Sitter space. In this case, $\Sigma$ would become
the equator of a constant time slice which has an $S^5$ geometry. The
EE would then be interpreted as measuring the entanglement of the CFT
between the two halves of this time slice. In fact, this will be the
entanglement entropy across the cosmological horizon of the de Sitter
geometry.\footnote{See \cite{cosmic,cthem} for a different
interpretation of this particular calculation.}

In fact, in the preceding example, one can readily see that the
universal contribution to the EE is simply proportional to the central
charge $A$. The reason being that the boundary geometry is conformally
flat and examining eqs.~\reef{Waldformula8z} and \reef{items9}, we see
that the expressions multiplying the $B_n$ are all proportional to the
Weyl tensor. Hence the latter contributions must all vanish in this
particular case. However, this example is instructive, as we learn that
to probe these terms in eq.~\reef{Waldformula8z}, we must choose the
boundary geometry to not be conformally flat. As a result, the bulk
geometry will not be simply pure AdS space. However, we will only need
to understand the details of the asymptotic geometry, as described in
appendix \ref{schwimmer}. In fact to determine the universal
contribution, \ie the logarithmic term, in the holographic EE with
$d=6$, we will have to carry this asymptotic expansion to second order.
For the boundary geometries chosen below, these expansions are
explicitly constructed in appendix \ref{curveb}.

We consider four different six-dimensional boundary geometries in the
following:  a) $R\times S^2\times S^3$, b) $R^3\times S^3$, c)
$R^2\times S^4$ and d) $S^3\times S^3$. Recall that the entangling
surface is a four-dimensional submanifold which we wish to choose in a
symmetric way so that there is a rotational symmetry in the transverse
two dimensions. For example, the backgrounds (a), (b) and (d) contain
an $S^3$ and we can choose $\Sigma$ to wrap a maximal $S^1$ in this
component of the geometry, as well as filling the other three
dimensions of the background. Similarly, in the geometry (c), $\Sigma$
can wrap a maximal $S^2$ inside the $S^4$ and also the $R^2$ component
of the boundary geometry. Alternatively, $\Sigma$ can wrap the entire
$S^4$ and sit at a point in the $R^2$. In the latter case, there is a
rotational symmetry in the plane $R^2$. Similarly, in the geometry (b),
we can also choose $\Sigma$ to wrap $R\times S^3$, which leaves a
rotational symmetry about the line $R$ in the $R^3$.

With the bulk metrics given in appendix \ref{curveb}, one must solve
for the asymptotic expansion \reef{braneform} of the bulk surface $m$
which minimizes the entropy functional \reef{Waldformula3z}. Because of
the rotational symmetry about $\Sigma$, the extrinsic curvatures vanish
on this surface. Then as can be seen from eq.~\reef{nextX}, this
vanishing implies that the expansion \reef{braneform} could only begin
at second order in $\tau$. However, we have also verified that in fact
for all of our examples below the second order term also vanishes
leaving
 \be
X^i(y^a,\tau) = \Xz{}^i(y^a) + \mathcal{O}(\tau^3)\,.
 \ee
The universal contribution in the holographic EE can be evaluated by
plugging these expressions back into \reef{Waldformula3z} and
extracting the logarithmic divergence.
%The corresponding AdS metric in Lovelock theories is obtained by solving
%up to second order in the FG expansion near the AdS boundary, which
%is then substituted into \ref{Waldformula3z}. The results are detailed
%in (\ref{schwimmer}). Taking the same
%gauge choice as in \cite{adam1} on the defect action, the expansion
%of the brane position near the AdS boundary actually yields
%\be
%X(x,\tau) = \Xz(x) + \mathcal{O}(\tau^3),
%\ee
%due to the vanishing boundary extrinsic curvature.
We now present the results of our holographic calculations and of the
CFT analysis \reef{Waldformula8z} for these universal
contributions for the various geometries:\\
 \noindent{\bf a) $R\times S^2\times S^3$ with $\Sigma=R\times S^2\times S^1$:}
 \bea S_{\mt{JM}}&=&\frac{3\pi
V_\Sigma}{100 R_1^4 R_2^4}\,\frac{\tilde{L}^5}{36\lp^5}
\bigg( 8 R_1^2 R_2^2 (3-13\fin\lambda - 30\fin^2\mu)\nonumber \\
&&\ -3R_1^4 (7-17\fin\lambda -30\fin^2\mu) +12 R_2^4 (3-13\fin\lambda -
30\fin^2\mu))\bigg)\log(\ell/ \delta)
\,,  \\
S_\mt{EE}&=&-\frac{3\pi V_\Sigma}{100 R_1^4 R_2^4} \bigg(B_1 (17 R_1^4 +
72 R_1^2 R_2^2 + 108 R_2^4)-4 B_2 (13 R_1^4 +
8 R_1^2 R_2^2 + 12 R_2^4) \nonumber \\
   &&\qquad+
   16 B_3 (17 R_1^4 + 32 R_1^2 R_2^2 + 48 R_2^4)\bigg)\log(\ell/ \delta)\,.
 \eea
where $R_1$ and $R_2$ are the radii of curvature for the $S^3$ and
$S^2$, respectively, and $\ell$ is some macroscopic scale of the CFT
geometry. Further $V_\Sigma$ is the volume of the entangling
surface, \ie for $\Sigma=R\times S^2\times S^1$, $V_\Sigma=8\pi^2R_1
R_2^2H$ where $H$ is a regulator length along the $R$ factor. In fact,
we do not explicitly need to evaluate $V_\Sigma$ to compare the two
expressions above.  Using the holographic expression \reef{center6} for
the four central charges, we find $S_{\mt{JM}}=S_\mt{EE}$. We present
the remaining results more briefly.\\
 \noindent{\bf b) $R^3\times S^3$ with $\Sigma=R^3\times
S^1$:}
 \bea S_{\mt{JM}} &=&
 \frac{\pi V_\Sigma}{25
R_1^4}\,\frac{\tilde{L}^5}{4\lp^5}\,
\left(3-13\fin\lambda -30 \fin^2\mu\right)\log(\ell/ \delta)\,,\\
S_{\mt{EE}}&=& - \frac{9\pi V_\Sigma}{25 R_1^4}\, \left(9 B_1 - 4 B_2 +
64 B_3\right)\log(\ell/ \delta)\,.
 \eea
where $R_1$ is the radius of the $S^3$.\\
 \noindent{\bf b') $R^3\times S^3$ with $\Sigma=R\times
S^3$:}
 \bea S_{\mt{JM}} &=&
 \frac{\pi V_\Sigma}{25
R_1^4}\,\frac{\tilde{L}^5}{4\lp^5}\,
\left(3-13\fin\lambda -30 \fin^2\mu\right)\log(\ell/ \delta)\,,\\
S_{\mt{EE}}&=&  -\frac{9\pi V_\Sigma}{25 R_1^4}\, \left(9 B_1 - 4 B_2 +
64 B_3\right)\log(\ell/ \delta)\,.
 \eea

 \noindent{\bf c) $R^2\times S^4$ with $\Sigma=R^2\times S^2$:}
 \bea
 S_{\mt{JM}}&=&-\frac{3\pi V_\Sigma}{400R_1^4}\,\frac{\tilde{L}^5}{\lp^5}
\left(9-19\fin\lambda-30\fin^2\mu\right)\log(\ell/ \delta)\,,\\
S_{\mt{EE}}&=& -\frac{3\pi V_\Sigma}{100 R_1^4}\,
\left(17 B_1 - 52 B_2 + 912
B_3\right)\log(\ell/ \delta)\,.
 \eea

 \noindent{\bf c') $R^2\times S^4$ with $\Sigma={\rm pt.}\times S^4$:}
 \bea
S_{\mt{JM}}&=&\frac{\pi V_\Sigma}{50
R_1^4}\,\frac{\tilde{L}^5}{4\lp^5}
\left(99-389\fin\lambda-2070\fin^2\mu\right)\log(\ell/ \delta)\,,\\
S_{\mt{EE}}&=& \frac{9\pi V_\Sigma}{50 R_1^4}\,
\left(\frac{25}{3\pi^3} A - 17 B_1 + 52 B_2 - 592
B_3\right)\log(\ell/ \delta)\,.
 \eea
where $R_1$ is the radius of the $S^4$.\\
 \noindent{\bf d) $S^3\times S^3$ with $\Sigma=S^1\times S^3$:}
 \bea
S_{\mt{JM}} &=& \pi V_\Sigma\, \frac{(R_1^2 + R_2^2)^2}{25 R_1^4
R_2^4} \,\frac{\tilde{L}^5}{4\lp^5}\left( 3-13\fin\lambda -30 \fin^2\mu
\right)\,\log(\ell/ \delta)\,, \\
S_\mt{EE}&=& -\pi V_\Sigma\, \frac{(R_1^2 + R_2^2)^2}{25 R_1^4 R_2^4} \,
9 (9 B_1 - 4 B_2 + 64 B_3)\,\log(\ell/ \delta)\,,
 \eea
where $R_1$ is the radius $S^3$ wrapped by the $S^1$ and $R_2$ is the
radius of the other $S^3$.

Again we do not need to explicitly specify $V_\Sigma$ in these
expressions to make the comparison of $S_\mt{JM}$ and $S_\mt{EE}$. In
every case, the holographic result $S_\mt{JM}$ is in complete agreement
with $S_\mt{EE}$ given by the CFT formula \reef{Waldformula8z}, when we
substitute in the holographic expressions for the central charges
\reef{center6}. Further note that with these tests, we have probed all
of the individual terms appearing in the CFT result, \ie all three
$B_n$, as well as $A$, appear in the expressions above. Hence once
again, we have found strong indications that $S_\mt{JM}$ is the correct
surface functional to extend the standard definition \reef{define} of
holographic EE to Lovelock gravity in the bulk.

\subsection{Entangling surfaces without rotational symmetry}
\label{broken}

As we noted above, the restriction for the CFT analysis in section
\ref{EEcft} is commonly stated as demanding that the extrinsic
curvature of $\Sigma$ should vanish, \eg \cite{solo}, rather than
requiring a rotational symmetry around this symmetry. Hence here we
examine the holographic EE for such surfaces, namely with zero
extrinsic curvature but without a rotational symmetry in the transverse
space. In particular, we focus on the first three backgrounds above,
where we can think of the geometry as $R_t \times \M_{d-1}$. Then,
because of the simple product form of the geometry, if the entangling
surface $\Sigma$ lies in the `spatial' geometry $\M_{d-1}$, the
extrinsic curvature associated with the normal vector in the time
direction vanishes, \ie $K^{\hat{t}}_{ab}=0$. At the same time, in our
examples, the `spatial' geometry contains various sphere components
$S^n$. If the entangling surface is chosen to wrap a maximal $S^{n-1}$
within one of these, we have ensured the vanishing of the remaining
extrinsic curvatures associated with a normal vector within $\M_{d-1}$.
Of course, as is clear in this construction, there will also be no
rotational symmetry around $\Sigma$. Below, we present the results for
the universal contribution to the EE from our holographic calculations
using eq.~\reef{Waldformula3z} and from the CFT analysis
\reef{Waldformula8z} in these geometries:\\
\noindent{\bf a) $R\times S^2\times S^3$ with $\Sigma=S^1 \times S^3$:}
 \bea
S_{\mt{JM}}&=& - \frac{\pi V_\Sigma}{1600 R_1^4 R_2^4
}\,\frac{\tilde{L}^5}{\lp^5}\,
  \bigg[3  (1+ 4\fin \lambda+15\fin^2\mu)R_1^4 -
    16  (3-13\fin\lambda-30\fin^2\mu)R_2^4\,,\nonumber  \\
 &&\quad -  4  (33-103\fin\lambda-210\fin^2\mu) R_1^2 R_2^2
 \bigg]\log(\ell/ \delta)\,, \\
S_{\mt{EE}}&=&\frac{3\pi V_\Sigma}{200 R_1^4 R_2^4}\,
 \bigg[(51 B_1 - 156 B_2 - 304 B_3)R_1^4 - 24 (9 B_1 - 4 B_2 + 64 B_3)R_2^4 \nonumber \\
&&\qquad - 6 (19 B_1 + 36 B_2 - 16 B_3)R_1^2R_2^2\bigg]\log(\ell/ \delta)\,,\\
\Delta S &=& S_{\mt{JM}}-S_{\mt{EE}}=\frac{\pi V_\Sigma}{64R_2^4}\frac{\tilde{L}^5}{\lp^5}(1-2\fin \lambda - 3\fin^2\mu)\log(\ell/ \delta)\,,
 \eea
where $R_1$ and $R_2$ are the radii of the $S^3$ and $S^2$,
respectively. Note that $\Sigma$ fills the entire $S^3$ and wraps a
maximal $S^1$ within the $S^2$ component. As before, the volume
$V_\Sigma$ is not needed to compare the two results. In the last line
above, we have substituted the holographic expression for the central
charges \reef{center6} into $S_{\mt{EE}}$ and we can see that there is
a discrepancy between the two results. Note, however, that the
difference $\Delta S$ is proportional to $B_3$ in eq.~\reef{center6}.
We present the remaining results more briefly.\\
\noindent{\bf a') $R\times S^2\times S^3$ with $\Sigma=S^2\times S^2$:}
 \bea
S_{\mt{JM}}&=&-\frac{\pi V_\Sigma}{1200 R_1^4 R_2^4
}\,\frac{\tilde{L}^5}{\lp^5}\,
    \bigg[3  (7-17\fin\lambda-30\fin^2\mu)R_1^4 +
      (39-134\fin\lambda-285\fin^2\mu)R_2^4 \nonumber \\
&&\quad-2  (87-337\fin\lambda-1950\fin^2\mu)R_1^2 R_2^2\bigg]\log(\ell/ \delta)\,,\\
S_{\mt{EE}}&=& -\frac{\pi V_\Sigma}{100 R_1^4 R_2^4} \bigg[3 (17 B_1 -
52 B_2 + 272 B_3) R_1^4 -8 (27 B_1 - 12 B_2 + 32 B_3) R_2^4 \nonumber \\
&&  \qquad - 6 (\frac{25}{3\pi^3} A + 19 B_1 + 36 B_2 - 336 B_3) R_1^2
R_2^2 \bigg]\log(\ell/ \delta)\,,\\
\Delta S&=&S_{\mt{JM}}-S_{\mt{EE}}=\frac{\pi V_\Sigma}{48R_1^4}\frac{\tilde{L}^5}{\lp^5}(1-2\fin \lambda - 3\fin^2\mu)\log(\ell/ \delta)\,,
 \eea
where again $R_1$ and $R_2$ are the radii of the $S^3$ and $S^2$,
respectively.\\
\noindent{\bf b) $R^3\times S^3$ with $\Sigma=R^2\times S^2$:}
 \bea
S_{\mt{JM}}&=& - \frac{\pi V_\Sigma}{1200
R_1^4}\,\frac{\tilde{L}^5}{\lp^5}\,
\left(39-134\fin\lambda-285\fin^2\mu\right)\log(\ell/ \delta)\,, \\
S_{\mt{EE}} &=& \frac{2\pi V_\Sigma}{25 R_1^4}\,\left(27 B_1 - 12 B_2
+ 32 B_3 \right)\log(\ell/ \delta)\,,\\
\Delta S&=&S_{\mt{JM}}-S_{\mt{EE}}=\frac{\pi V_\Sigma}{48R_1^4}\frac{\tilde{L}^5}{\lp^5}(1-2\fin \lambda - 3\fin^2\mu)\log(\ell/ \delta)\,, \eea
where $R_1$ is the radius of the $S^3$.\\
\noindent{\bf c) $R^2\times S^4$ with $\Sigma=R^1\times S^3$:}
 \bea
S_{\mt{JM}}&=&\frac{9\pi V_\Sigma}{1600
R_4^4}\,\frac{\tilde{L}^5}{\lp^5}\,
 (13-28\fin\lambda-45\fin^2\mu) \log(\ell/ \delta)\,,\\
S_{\mt{EE}}&=&\frac{3\pi V_\Sigma}{200 R_4^4}\, (51 B_1 - 156 B_2 +
1616 B_3)\log(\ell/ \delta)\,,\\
\Delta S&=&S_{\mt{JM}}-S_{\mt{EE}}=\frac{\pi V_\Sigma}{64R_4^4}\frac{\tilde{L}^5}{\lp^5}(1-2\fin \lambda - 3\fin^2\mu)\log(\ell/ \delta)\,,  \eea
where $R_1$ is the radius of the $S^4$.\\
%
%\underline{Case VI: Wrapping $S^5$ in $R^1 \times S^5 $}
%\bea
%S_{JM}&=& -2\pi V_{S^5}\frac{\tilde{L}^5}{\lp^5}\frac{3-10\fin\lambda -15\fin^2\mu}{8 R_5^4 \fin}\,,\\
%S_{\mt{EE}}&=&-V_{S^5} \frac{3 A}{2 \pi^2 R_3^4}\,.
%\eea
Hence in all these examples, there is a mismatch between the
holographic result found using eq.~\reef{Waldformula3z} and the CFT
result evaluated with eq.~\reef{Waldformula8z}. A clue as to the nature
of this mismatch comes from observing that in each of the above cases,
the difference $\Delta S$ is proportional to the holographic expression
for $B_3$, given in eq.~\reef{center6}. Further, we note that the
mismatch persists even in the limit of Einstein gravity, where we set
$\lambda=\mu=0$ which sends $\fin \to 1$.

We are specifically probing the EE in geometries where we know the
derivation sketched in section \ref{EEcft} is not valid and hence the
results for $S_\mt{EE}$ are suspect. Hence, in proceeding to examine
this mismatch, our working hypothesis will be that in fact the
holographic results above are actually the correct ones.

To understand the nature of the mismatch between the two calculations
even at vanishing extrinsic curvatures, we resort once more to the
powerful techniques developed in \cite{adam} to extract analytically
the holographic entanglement entropy for general boundary geometries
and submanifolds, as shown in section \ref{sch4d}. Above, we noted that
mismatch occurs even when the bulk theory is Einstein gravity where the
holographic EE is simply given by eq.~\reef{define}. To simplify the
calculations here then, we will restrict our attention to Einstein
gravity. However, we will be able to extend our results to remove the
discrepancies above for the cubic Lovelock theory.

In case of Einstein gravity, the holographic entanglement entropy is
simply the area of the minimal surface, whose expansion near the
boundary is readily obtained, analogous to the computations detailed in
section \ref{sch4d}. The only difference in six dimensions, however, is
that the measure, \ie $\sqrt{h}$ in the area integral, begins at order
$\tau^{-d/2}=\tau^{-3}$. Hence we extend our expansions to order
$\tau^2$ in order to extract the log-divergent term. This is a tedious
task, even without the complication of higher derivative corrections in
$S_\mt{JM}$. Hence to simplify our calculations further but yet keep
enough generality to encompass all of the geometries that are commonly
considered, \eg entangling surfaces as proposed above, we make the
following  assumptions about the geometry of the entangling
surface $\Sigma$ in the boundary metric:
 \be K^{\hi}_{a b} =0\,,
 \qquad R_{m n p r}\, n^{\hi\,m}\,n^{\hj\,n}\,n^{\hat{k}\,p}
 \,t^{r}_a=0\,,
 \labell{assumptions}
 \ee
In terms of the holographic construction, the tangent vectors are given
by $t^m_a= \partial_a\overset{\scriptscriptstyle{(0)}}{X^m}$ and $R_{m
n pr}$ is the boundary curvature tensors constructed from
$\overset{\scriptscriptstyle{(0)}}{g}_{ij}$. Together the above
assumptions imply that on $\Sigma$,
 \be
R_{m n}\,n^{\hi\,m}\,t^{n}_a =0\,. \ee

With these assumptions, the log term in the holographic entanglement
entropy with Einstein gravity is given by
 \be
S =2\pi\log \delta\,\frac{\tilde{L}^5}{\lp^5}
\int_{\Sigma}d^4x\sqrt{h}\left[\frac{1}{2}h^{ij} \gs_{ij} +
\frac{1}{8}(h^{ij} \go_{ij})^2 -\frac{1}{4}\go_{ij}\, h^{jk} \go_{kl}\,
h^{li}\right]\labell{gogo}
 \ee
where
 \be
h^{i j} = \hz^{a b}\,\partial_a\! \Xz^i\, \partial_b\! \Xz^{j}
 \ee
is the tangential projector with respect to $\Sigma$. Alternatively, we
can express this tensor as $h^{ij}=
\overset{\scriptscriptstyle{(0)}}{g}{}^{ij}-n^{\hi\,i}n^{\hi\,j}$.
Explicit expressions for $\overset{\scriptscriptstyle{(2)}}{g}$ and
$\overset{\scriptscriptstyle{(1)}}{g}$ are given in
eq.~(\ref{metricexpand}). One can check that, subject to our
assumptions, the above expression \reef{gogo} is conformally invariant
if we restrict to transformations to be independent of the transverse
coordinates. Note that each of the terms in CFT result
(\ref{Waldformula8z}) is also conformally invariant in the same sense.

Now one can compute the difference between the holographic result
\reef{gogo} and the expected CFT result (\ref{Waldformula8z}) for
Einstein gravity. Clearly this difference has to be some conformal
invariant that vanishes when evaluated on defects that preserve the
rotational symmetry in the transverse space. It is interesting that in
the difference, terms with explicit covariant derivatives of the
curvatures exactly cancel. In light of conformal invariance, the result
can be arranged into the following compact form:
 \bea
\Delta S = 4\pi B_3\log \de\, \int_{\Sigma}d^4x\sqrt{h}(&& C_{m n}{}^{r
s }C^{m n k l} \tilde{g}^\perp_{s l}\tilde{g}^\perp_{r k} -C_{m n
r}{}^s C^{m n r l}\tilde{g}^\perp_{s l}
 \labell{discrepancy}  \\
&&+ 2 C_{m}{}^n{}_r{}^s C^{mkrl} \tilde{g}^\perp_{n
s}\tilde{g}^\perp_{k l} - 2C_{m}{}^n{}_r{}^s C^{mkrl}
\tilde{g}^\perp_{n l}\tilde{g}^\perp_{k s})\,,
 \nonumber
 \eea
where $\tilde{g}^\perp_{ij}= n^\hi_i n^\hi_j$ is the metric in the
transverse space to $\Sigma$. We have written the coefficient as
$B_3=\frac{\tilde{L}^5}{384\lp^5 }$, even though in Einstein gravity
the central charges can not really be distinguished since they are all
proportional to $\tilde{L}^5/\lp^5$. However, our previous results
suggest that $\Delta S$ will be proportional to $B_3$ in a more general
context.

In the presence of a rotational symmetry in the transverse space, any
tensors with an odd number of projectors into the transverse space must
vanish. Further, the rotational symmetry guarantees that the extrinsic
curvatures all vanish. It is fortuitous then that these observations
are consistent with our assumptions above in eq.~(\ref{assumptions}).
For a tensor that carries only two transverse indices, its symmetric
part, should be proportional to the transverse metric
$\tilde{g}^\perp$, and the anti-symmetric part, the volume form
$\tilde{\eps}$. Hence we can write
 \be
C_{a n_i b n_j}= H^S_{a b}\,\tilde{g}^\perp_{n_i n_j} + H^A_{a b}\,
\tilde{\eps}_{n_i n_j}\,,\labell{gogo9}
 \ee
where $H^{(A)S}_{a b}$ is some (anti)symmetric tensor, as required by
the symmetry of the Weyl tensor, with tangential indices along
$\Sigma$. Eq.~\reef{gogo9} implies, from the cyclic permutation
property of the Weyl tensor, that
 \be
C_{a b n_i n_j}=2H^A_{a b}\,\epsilon^\perp_{n_i n_j}\, .
 \ee
Substituting these expressions into eq.~\reef{discrepancy}, and using
the fact that the transverse space is two-dimensional,\footnote{The latter
ensures that the components containing only transverse indices, \ie
$C_{n_1n_2n_3n_4}$, cancel altogether.} one can verify
that $\Delta S$ vanishes identically in the presence of rotational
symmetry.

We have derived $\Delta S$ from a holographic analysis with Einstein
gravity, however, our conjecture is that this correction term should be
included in general. That is, we propose that the correct result for
the universal EE in a six-dimensional CFT is given by the sum of
$S_\mt{EE}$ in eq.~\reef{Waldformula8z} and $\Delta S$ in
eq.~\reef{discrepancy} --- of course, this should only be the correct
result when the extrinsic curvature of $\Sigma$ vanishes. We can test
this conjecture with the holographic results of the cubic Lovelock
theory above. In this comparison, we have verified that in fact $\Delta
S$ in eq.~\reef{discrepancy} precisely matches the difference $\Delta
S$ in each of these four examples, which then seems to be strong
evidence in favour of our proposal. Again, this proposal only applies
when the extrinsic curvature of $\Sigma$ vanishes. When the extrinsic
curvature is also nonvanishing, there should be many more correction
terms. One could try to determine these terms following the analysis of
\cite{solo}. First, one constructs all possible conformal invariants
involving four derivatives, constructed with the extrinsic curvatures
(and the Weyl tensor). Next, one assembles these invariants with
arbitrary coefficients in a trial expression which would be added to
$S_\mt{EE}+\Delta S$. Then, one tries to fix the coefficients by
testing this CFT result for a variety of entangling surfaces and
background geometries against the holographic result for the cubic
Lovelock theory.

\section{General gravity actions} \label{general}

The two previous sections show quite convincingly that holographic EE
in Lovelock gravity is again calculated by minimizing a surface
functional and that the appropriate functional is given by the
expression $S_\mt{JM}$ in eq.~\reef{Waldformula3}, originally derived
to evaluate black hole entropy in this theory. Now a natural question
is whether this success teaches us any lessons for a more general
gravity action in the bulk. Unfortunately, the lessons may be limited.
For example, the derivation \cite{ted1} of black hole entropy leading
to eq.~\reef{Waldformula3} relies on a Hamiltonian formulation which is
not readily extended beyond Lovelock gravity, \ie a generic higher
curvature theory will not have second order equations of motion.
However, working with Lovelock gravity has certainly provided a great
deal of experience with regards to holographic EE and so here we will
try to apply this experience to a more general gravity theory.

As an interesting test case, we focus on a general curvature-squared
action with $d=4$
 \beq
I = \frac{1}{2\lp^3} \int \mathrm{d}^5x \, \sqrt{-g}\, \left[
\frac{12}{L^2} + R + L^2\left(\la_1\, R_{\mu\nu\rho\sigma}
R^{\mu\nu\rho\sigma} +\la_2\,R_{\mu\nu}R^{\mu\nu}+\la_3\,R^2
\right)\right]\ .
 \labell{actorg}
 \eeq
In examining this theory, a typical approach would be that the
curvature-squared terms appear as the first few corrections in a
perturbative string expansion, \eg \cite{alex,alex2}. In this context,
we would have small couplings, \ie $\lambda_{1,2,3}\ll1$, and we would
only calculate to leading order in any of these parameters. If one
attempted to work with the full nonlinear theory (and finite
couplings), one encounters the typical problems. For example, the
generic action leads to fourth order equations of motion which produces
ghost-like excitations in the gravitational theory and from a
holographic perspective, this corresponds to producing nonunitary
operators to the boundary CFT \cite{cosmic}. Of course, if we tune the
couplings as
 \be
 \lambda_1=\lambda_3=\lambda/2\quad{\rm and}\quad
 \lambda_2=-2\lambda\,,
 \labell{tuneGB}
 \ee
this action \reef{actorg} coincides with Gauss-Bonnet (GB) gravity
\reef{GBact2} and this problem of higher derivatives is evaded. In the
following, we will examine the theory primarily from the perturbative
perspective but we will not set the couplings to be small in our
analysis as this will allow us consider the case of GB gravity further,
as well.

The AdS$_5$ vacua now have curvature $\tL^2=L^2/\fin$ where
\cite{highc,cosmic}
 \be
1=\fin-\frac23 \fin^2\left(\la_1+2\la_2+10\la_3\right)\,.
 \labell{rootg}
 \ee
If we are treating the higher curvature terms perturbatively, \ie
$\la_{1,2,3}\ll1$, this then would yield
 \be
\fin=1+\frac23\left(\la_1+2\la_2+10\la_3\right)+\cdots\,.
 \labell{rootg2}
 \ee
Of course, with the GB couplings \reef{tuneGB}, the expression
\reef{rootg} above agrees with eq.~\reef{roots} for GB gravity.
Further, ref.~\cite{highc} evaluated the holographic trace anomaly for
this general action \reef{actorg} and one finds
 \be
c= \pi^2\frac{\tL^3}{\lp^3}\left(1+4\fin\left(\la_1-2\la_2-10\la_3
\right)\right)\ ,
 \qquad
a=\pi^2\frac{\tL^3}{\lp^3}\left(1-4\fin\left(\la_1+2\la_2+10\la_3
\right)  \right)\ .
 \labell{central3}
 \ee
Again, with eq.~\reef{tuneGB}, this agrees with the result \reef{ccaa}
determined for GB gravity.

Now we would like to consider holographic EE in the presence of these
general curvature-squared interactions. As a first approximation, we
take our surface functional to be the Wald formula.\footnote{To
simplify the notation slightly, we will assume that we have Wick
rotated to Euclidean signature in the following discussion.} Upon
evaluating eq.~\reef{Waldformula} for the above action \reef{actorg},
the result can be written as
 \beqa
S_\mt{W} &=& \frac{2 \pi}{\lp^{3}} \int_m d^{3}x\sqrt{h}\,\left[ 1+
L^2\left((2\la_1+\la_2+2\la_3)\,R^{\mu\nu\rho\sigma}
\tilde{g}^{\perp}_{\mu\rho}\tilde{g}^\perp_{\nu\sigma}
\right.\right.\nonumber\\
&&\qquad\qquad\quad\left.\left. +(\la_2+4\la_3)\,R^{\mu \al\nu
\bt}\tilde{g}^{\perp}_{\mu\nu} h_{\al\bt}
+2\la_3\,R^{\al\bt\ga\de}h_{\al\ga}h_{\bt\de} \right) \right]\,,
 \labell{Waldformula55}
 \eeqa
where $\tilde{g}^\perp_{\mu\nu}$ and $h_{\al\bt}$ are the transverse
and induced metrics for the surface $m$, respectively --- here, we are
applying the notation introduced in section \ref{eed4} to the bulk
surface $m$. Note that with the GB couplings \reef{tuneGB}, the
coefficients of the first two curvature terms above vanish. In general,
if we apply the expressions for the central charges (\ref{central3}),
then this Wald expression (\ref{Waldformula55}) will produce results
for the EE which agree with those coming from the CFT analysis, \ie
eq.~(\ref{Waldformula8yz}), but only for entangling surfaces on which
the extrinsic curvature vanishes. Applying the techniques of section
\ref{sch4d}, we find the holographic EE contains a logarithmic term of
the form
 \be
S_\mt{W} = \frac{\log(\ell/\delta)}{2\pi}\int_{\Sigma}d^2x \sqrt{h}
\left[c \,C_{abcd}h^{ac} h^{bd} -a \left(\R + (K^\hi_a{}^b
K^\hi_b{}^a-\frac12 K^\hi_a{}^a K^\hi_b{}^{\,b})\right)\right] \,.
\labell{gwald}
 \ee
Here, the Weyl tensor corresponds to that evaluated for the boundary
metric $\overset{\scriptscriptstyle{(0)}}g_{ij}$ while intrinsic
curvature $\R$ and the extrinsic curvatures are evaluated on the
entangling surface $\Sigma$, again embedded in the boundary geometry
$\overset{\scriptscriptstyle{(0)}}g_{ij}$. Note that this expression
\reef{gwald} is composed of three independent conformal invariants. Now
to fix eq.~(\ref{gwald}) to agree with eq.~\reef{Waldformula8yz} from
the pure CFT analysis, we should presumably add extra terms to the
surface functional. It is reasonable to assume that these new terms
should be covariant and contain only two derivatives, but be
independent of the terms already appearing in eq.~\reef{Waldformula55}.
It seems then that the only natural geometric terms will be constructed
from the extrinsic curvature of the bulk surface
$m$,\footnote{Following our notation in footnote \ref{footy}, the
extrinsic curvatures on $\Sigma$ and $m$ are distinguished by the type
of indices, \ie Latin and Greek indices for $\Sigma$ and $m$,
respectively.} which we will denote $K^\hi_{\al\bt}$. There are two
independent terms and so we write
 \be
\delta S = \frac{2 \pi L^2}{\lp^{3}} \int_m d^{3}x\sqrt{h}\, \left(s_1\,
K^\hi_\al{}^\bt K^\hi_\bt{}^\al+s_2\, K^\hi_\al{}^\al
K^\hi_\bt{}^{\,\bt}\right) \,.
 \labell{Waldfix0}
 \ee
Now we want to fix the value for constants $s_1$ and $s_2$ so that
there is an additional contribution to the logarithmic term with
precisely the form
 \be
\delta S = (a-c) \frac{\log(\ell/\delta)}{2\pi}\int_{\Sigma}d^2x
\sqrt{h} \left(K^\hi_a{}^b K^\hi_b{}^a-\frac12 K^\hi_a{}^a
K^\hi_b{}^{\,b}\right)\,.
 \labell{fixx2}
 \ee
With some further analysis, we find the desired result requires
 \be s_1 = -2\lambda_1\,,\labell{Waldfix} \ee
while $s_2$ remains undetermined. That is, the term
$(K^\hi_\al{}^\al)^2$ in eq.~\reef{Waldfix0} only contributes to
regular terms in the entanglement entropy and it does not contribute to
the universal logarithmic term (or to the nonuniversal divergent
terms). In a perturbative framework, this ambiguity cannot be resolved.
At zeroth order in the $\la_i$, the entropy is simply given by an
extremal surface in the AdS bulk, which then necessarily satisfies
$K^\hi_\al{}^\al=0$, \eg see \cite{zeroextrinc}. Assuming that
$s_2=O(\la_i)$, then this term would only begin to contribute at order
$\la_i^3$. Hence it simply does not effect the holographic EE at the
linear order, which is the order of validity of the present
calculations in the perturbative expansion.

Note that in the perturbative framework where $\la_i\ll1$, the
couplings $\la_2$ and $\la_3$ can be varied (and even be set to zero)
by field redefinitions (\eg $\delta g_{\mu\nu}=\al_1 R_{\mu\nu} +\al_2
R\, g_{\mu\nu}$) but these changes should not change any physical
quantities, \eg see \cite{alex}. The effect of such field redefinitions
on the holographic EE may seem a bit mysterious given
eq.~\reef{Waldformula55}. However, it is perhaps clearer when we note
that the $\la_2$ and $\la_3$ contributions there can be rewritten in
terms of just $R_{\mu\nu}$ and $R$ --- see eq.~\reef{Waldformula55x}
below. In any event, it is reassuring that our results for the
universal contribution to the holographic EE in eqs.~\reef{gwald} and
\reef{fixx2} can be written entirely in terms of the central charges
(and the geometry of the entangling surface $\Sigma$) in the CFT. From
this perspective, it is also interesting that the coefficient $s_1$ is
fixed in eq.~\reef{Waldfix} in terms of only $\la_1$, the single
coupling whose value is not subject to field redefinitions.

With such field redefinitions, we could always tune the
curvature-squared couplings to
 \be
\la_1=\la\,,\qquad \la_2=-\frac43\la\,,\qquad \la_3=\frac16\la\,.
 \labell{tuneW2}
 \ee
In this case, the higher curvature terms in eq.~\reef{actorg} combine
as $L^2\lambda\,C_{\mu\nu\rho\sigma}C^{\mu\nu\rho\sigma}$, \ie the Weyl
tensor squared. Further the curvature terms in the Wald contribution
\reef{Waldformula55} to the holographic EE would be proportional to
$C^{\mu\nu\rho\sigma}\tilde\veps_{\mu\nu}\tilde\veps_{\rho\sigma}$. In
this case, this term in Wald entropy will simply not contribute to any
calculation of holographic EE in pure AdS$_5$ and the entire $\la$
contribution will come from the correction term \reef{Waldfix0}.
However, this term proportional to the Weyl tensor can still contribute
here in more general backgrounds, such as considered in section
\ref{sch4d}. In fact, following the analysis there, one finds that in a
general background, the components of the bulk Weyl tensor scale as
$C^{\mu\nu}{}_{\rho\sigma}\sim\rho$ near the boundary and hence this
term will typically contribute to the universal term in the holographic
EE.

There is, of course, a well-known higher curvature term in string
theory which is quartic in the Weyl tensor \cite{curv4}. In this case,
the interaction would produce a Wald contribution to the holographic EE
which is proportional to the Weyl tensor cubed and so again this
contribution would vanish in pure AdS$_5$ --- implicitly, we will focus
on $d=4$ here. Further following the analysis of section \ref{sch4d},
this $C^3$ term would generically vanish at least as fast as $\rho^3$
near the asymptotic boundary and hence it would never be able to
contribute to the universal EE term. This result is, in fact,
essentially required by the consistency of the holographic framework.
From the perspective of the boundary theory, these $C^4$ terms
introduce corrections of order $1/\lambda^{3/2}$ and
$\lambda^{1/2}/N_c^2$ \cite{univres} and in particular then, these
corrections depend on the 't Hooft coupling $\lambda$. However, the
analysis of section \ref{EEcft} indicates that the universal
contribution to the EE should be proportional to the central charges
$a$ and $c$ and in a four-dimensional superconformal gauge theory, it
is known that the central charges are independent of the gauge coupling
\cite{nochange}. Therefore this universal term can not receive any
$\lambda$-dependent corrections.\footnote{We might add that for $N=4$
super-Yang-Mills theory in the free field limit, numerical calculations
\cite{numerical} of the EE for a sphere embedded in flat space
explicitly confirm that the universal contribution matches the strong
coupling result and so also confirms this independence of the gauge
coupling.} This is certainly in accord with our conclusion above that
the Wald contribution coming from the $C^4$ interaction does not
contribute to the universal term. However, just as in our analysis of
the curvature-squared interactions above, we expect that the correct
functional for the holographic EE will receive corrections beyond this
Wald contribution. Hence one restriction on these corrections is that
they can not contribute to the universal EE term for any background or
for any entangling surface. In fact, it seems that this constraint is
easily satisfied. A preliminary analysis suggests that covariant terms
of the form $C^2 K^2$, $C K^4$ or $K^6$ all vanish as fast as $C^3$
near the boundary, \ie at least as fast as $\rho^3$. Hence none of
these potential contributions to the surface functional would affect
the universal EE term. We emphasize that all of these terms, as well as
the original $C^4$ interaction, would still make finite contributions
to the EE, \eg in a thermal state, the temperature dependence of the EE
would receive finite $\lambda$ corrections.

\subsection{An ambiguity in holographic EE?} \label{ambigx}

Let us return to considering the curvature squared theory \reef{actorg}
with the coefficients tuned as in eq.~\reef{tuneGB} to produce GB
gravity. Above we identified an ambiguity in the correction term
\reef{Waldfix0}, in that we did not fix the coefficient $s_2$. In the
perturbative framework, we showed this ambiguity would not affect the
results for the holographic EE since the corresponding contribution was
always higher order in $\lambda_i$. However, for GB gravity where the
couplings are kept finite, the story is more interesting.

Na\"ively, our expectation would be that the coefficients in the
correction term \reef{Waldfix0} should be fixed so that the Wald
expression \reef{Waldformula55} is converted to the expression
\reef{Waldformula3x} which was successfully tested in section
\ref{EE4}. This would, in fact, require that $s_1$ takes the value
given in eq.~\reef{Waldfix} but it would also require that
$s_2=-s_1=2\lambda_1$. Of course, the analysis above showed that this
coefficient is simply not fixed if we only demand that the holographic
entanglement entropy reproduce the correct logarithmic term. We note
that the latter was precisely the criterion against which we tested
eq.~\reef{Waldformula3} in sections \ref{EE4} and \ref{EE6}. Hence our
analysis there is actually incomplete, since we have shown here that
this leaves certain ambiguities in the definition of the surface
functional used to calculate the holographic EE.

Hence we must find another approach to fix this amibiguity. To produce
a well-defined variation problem, it is reasonable to require that the
equations of motion fixing the extremal surface in the bulk should
remain second order. Since the Wald part of the surface functional
\reef{Waldformula55} contains only projectors of the bulk curvatures
into the surface world-volume, they contribute only terms which are
second order in derivatives to the equations of motion. The only source
of higher derivative terms comes from  the correction term
\reef{Waldfix0}. Thus we would like to find a suitable ratio of the
coefficients, $s_{1},\,s_{2},$ such that any higher derivative terms in
the equations of motion cancel.

Since we are varying only the embedding of the surface $m$, we can
safely choose a convenient gauge for the background metric. We opt for
Riemannian normal coordinates so that the Christoffel symbols are set
to zero locally in the vicinity of any point on $m$. Of course,
derivatives of the connection will not vanish in general, but one can
show that these terms do not contain cubic or higher derivatives of the
embedding function. Therefore we can effectively consider a flat
Minkowski background, in which case, eq.~\reef{Waldfix0} simply reduces
to
 \be
\delta S = \frac{2 \pi L^2}{\lp^{3}} \int d^{3}x\sqrt{h}\, \left[
\tilde{g}^\perp_{\mu\nu}\partial_\al \partial_\bt X^{\mu}\,\partial_\ga
\partial_\de X^\nu\, \left(s_1\,h^{\al \ga}h^{\bt \de} +s_2 \, h^{\al \bt}
h^{\ga \de} \right)\right]\,. \labell{fourderz}
 \ee
The four-derivative terms in the equations of motion are then given by
 \be
\partial_\al \partial_\bt\left(\frac{\delta\, (\delta S)}{\partial_\al
\partial_\bt X^\mu}\right)\bigg\vert_{\textrm{4-derivative}} = \frac{4
\pi L^2}{\lp^{3}}(s_1+s_2)\, \sqrt{h}\, \tilde{g}^\perp_{\mu\nu}\,h^{\al
\bt}\,h^{\ga \de}\,
\partial_\al \partial_\bt \partial_\ga \partial_\de X^{\nu}\, .
 \labell{fourderx}
 \ee
This immediately singles out the special value
 \be s_2 = - s_1\,.\labell{twoderivative} \ee
Further we have checked that with this choice of the coefficients, the
three-derivative terms also vanish.

Hence, eq.~\reef{twoderivative} guarantees that the variational problem
produces only two-derivative equations. However, as noted above, this
constraint, together with eq.~\reef{Waldfix}, give precisely the
necessary coefficients to convert the Wald entropy functional
\reef{Waldformula55} to $S_{\mt{JM}}$, given in
eq.~\reef{Waldformula3x}. Hence we have uniquely determined $S_\mt{JM}$
as the correct surface functional in calculating holographic EE for GB
gravity with two criteria. First, the holographic entanglement entropy
must reproduce the correct logarithmic term and second, the variational
problem must be second order in derivatives. While we have not
investigated the latter criterion in detail for higher Lovelock
theories, we note that eq.~\reef{Waldformula3} is constructed with
extended Euler densities for the intrinsic surface geometry. Of course,
they have the same structure as the Lovelock action \reef{GBactg}
itself and so one expects that an analysis similar to that showing the
Lovelock equations are second order would show the variational problem
here is also second order. Hence we expect that the same two criteria
above will also uniquely select $S_\mt{JM}$ as the appropriate surface
functional to calculate holographic EE for the general Lovelock
theories, as well.

\section{Discussion} \label{discuss}

The present paper was an exploration of holographic entanglement
entropy for higher curvature gravity theories. We were naturally led
to consider a procedure of extremizing some surface functional, similar
to the original definition \reef{define} for Einstein gravity, in order
that the holographic EE satisfies subadditivity \reef{ssub}. The close
connection with black hole entropy suggests that the new functional
might simply be Wald's formula \reef{Waldformula}. However, one of our
results, in section \ref{not}, was that this prescription would fail to
provide the correct EE in general. This is unfortunate as it would have
given a simple prescription that could be applied quite generally to
any higher curvature theory of gravity.

Turning to the special case of Lovelock gravity \reef{GBactg}, we
considered an alternative expression \reef{Waldformula3}, which still
coincides with Wald's formula on the Killing horizon of a stationary
black hole. In sections \ref{EE4} and \ref{EE6}, we showed that
extremizing $S_\mt{JM}$ yields the correct universal EE contribution
for CFT's in $d=4$ and 6 with a variety of geometries. In fact, in
$d=4$, we showed that the holographic approach precisely reproduced the
general expression \reef{Waldformula8yz} for the universal contribution
for any smooth entangling surface. In $d=6$, we found a precise match
for various geometries where the background geometry was not
conformally flat and the entangling surface was chosen so that there
was a rotational symmetry around the surface.\footnote{Note that for
the geometries chosen for $d=6$, the extremal bulk surface has
vanishing extrinsic curvature and so on this surface
$S_\mt{W}=S_\mt{JM}$. We also note that the latter observation is not
in contradiction with the result in section \ref{not} that
$S_\mt{W}\propto A$ for any entangling surface because this only
applies for empty AdS space.} While our approach of testing $S_\mt{JM}$
focussed on even dimensions and on the vacuum of the boundary CFT, we
expect that the result is quite general. That is, for any Lovelock
theory in any dimension and in any asymptotically AdS geometry, the
holographic EE can be calculated by extremizing the $S_\mt{JM}$
functional \reef{Waldformula3} for surfaces homologous to the boundary
region of interest.

In section \ref{ambigx}, we found a potential ambiguity in our
prescription for Gauss-Bonnet gravity. In particular, a term
proportional to the square of the trace of the extrinsic curvature
could be added to $S_\mt{JM}$ with an arbitrary coefficient and still
leave unchanged the universal EE contribution. We emphasize that this
additional term would still modify the finite contributions to the
entanglement entropy. However, we argued that the coefficient of this
extra term must be set to zero in order to preserve the fact that the
variational problem in calculating the holographic EE is still second
order in derivatives. While our analysis here focused on Gauss-Bonnet
gravity in five bulk dimensions, it extends trivially to any spacetime
dimension. We also expect that similar ambiguities will arise for
higher Lovelock theories but that again requiring a second order
variational problem will uniquely select $S_\mt{JM}$ as the appropriate
surface functional.

The goal remains to determine a comprehensive prescription for
holographic EE which can be applied to general higher curvature
theories. So it is natural to ask whether our success in understanding
holographic EE in Lovelock gravity can teach us any lessons for a more
general gravity actions in the bulk. Unfortunately, it seems that the
lessons may be limited. It is reasonable to expect that the special
form of $S_\mt{JM}$ which only involves the intrinsic curvature of the
surface on which it is evaluated must be related to the topological
origin of the Lovelock theories. However, consider the following
analysis: In section \ref{general}, we considered a general curvature
squared action \reef{actorg} and it is clear that the final surface
functional $S_\mt{W} +\de S$ depends on more than just the intrinsic
geometry, if we examine eq.~\reef{Waldformula55}. However, we observe
that we can rewrite the expression as
 \beqa
S_\mt{HEE}&=&\left[S_\mt{W}+\delta S\right]_{s_2=-s_1=2\la_1}\nonumber\\
&=& \frac{2 \pi}{\lp^{3}} \int_m d^{3}x\sqrt{h}\,\left[ 1+
L^2\left(2\la_1\,\R-(4\la_1+\la_2)\,R^{\al\bt}h_{\al\bt}
\right.\right.\nonumber\\
&&\qquad\qquad\qquad\qquad\left.\left. +(2\la_1+\la_2+2\la_3)\,R\right)
\right]\,,
 \labell{Waldformula55x}
 \eeqa
where $\R$ denotes the intrinsic curvature scalar $m$. Producing this
final expression relied on a number of geometric identities, \eg the
fact that the Weyl tensor is traceless, but also fixing $s_2$ as in
section \ref{ambigx}. Of course, if we choose the couplings $\la_i$ as
in eq.~\reef{tuneGB}, corresponding to GB gravity, the coefficients of
the last two terms vanish and we recover $S_\mt{JM}$ again. On the
other hand, one might also consider this expression in a perturbative
framework (with $\la_i\ll1$) in which case we can substitute the
leading order gravitational equations into the above expression. That
is, with $R_{\mu\nu}=-4 g_{\mu\nu}/L^2+O(\la_i)$, eq.~\reef{Waldformula55x}
reduces to
 \beqa
S_\mt{HEE} &=& \frac{2 \pi}{\lp^{3}} \int_m d^{3}x\sqrt{h}\,\left[ 1+
8(\la_1-\la_2-5\la_3)+ 2L^2\,\la_1\,\R\right]+O(\la_i^2)
 \labell{Waldformula55z}
 \eeqa
Hence there is also a sense that, within the perturbative framework,
the functional determining the holographic EE only depends on the
intrinsic geometry of the surface. Of course, the final expression
would be slightly more complicated if the bulk theory coupled the
gravitational theory \reef{actorg} to various matter fields. Then it
appears that simplifying with the gravitational equations of motion
would introduce matter field terms into eq.~\reef{Waldformula55z}. In
any event, it seems that further explorations will be needed before a
comprehensive prescription for holographic entanglement entropy is
uncovered. It may be interesting to explore these issues with the
physically reasonable theories constructed in \cite{cosmic} with cubic
curvature interactions.

As discussed in section \ref{EEcft}, the universal term in the EE of a
CFT can be determined in terms of the central charges using an analysis
involving the trace anomaly \cite{rt2,solo} --- see also discussion in
\cite{cosmic}. The results of this analysis formed the basis of our
consistency tests for various prescriptions for holographic EE.
However, the CFT analysis can only be applied in situations where there
is a rotational symmetry in the transverse space about the entangling
surface $\Sigma$. Hence it is known that this analysis does not capture
any of the contributions involving the extrinsic curvature \cite{solo}.
However, in section \ref{broken}, we found however that there can also
be various terms just involving the bulk curvature which are also
missed in this analysis. The new terms \reef{discrepancy} which we
found there correct the standard result \reef{Waldformula8z} for $d=6$.
However, we expect that there will be similar corrections involving
only bulk curvatures for any $d\ge6$. It seems that this was not a
problem in $d=4$ simply because the low dimension limits the number of
conformal invariants \cite{solo}. Of course, there will also be a
variety of further corrections involving extrinsic curvatures to our
results in $d=6$ or for higher dimensions. In any event, our results
highlight the necessity of a rotational symmetry about the entangling
surface to apply the analysis of \cite{rt2,solo}. It is incorrect to
describe the necessary requirement as saying the extrinsic curvatures
must vanish, as is commonly done.

In the holographic framework, when the entangling surface $\Sigma$ has
a rotational symmetry boundary, this typically extends to a symmetry
about a bulk surface $m_{\ssc\Sigma}$. The latter then naturally
becomes the extremal surface in calculating the holographic EE. In such
a situation, it also appears that upon analytically continuing back to
Minkowski signature, the rotational symmetry will become a Killing
symmetry, but further that $m_{\ssc\Sigma}$ becomes the bifurcation of a
Killing horizon in the new Minkowski signature spacetime. That is, the
resulting bulk geometry has the structure of a black hole. One
obstruction to the latter may be if somehow a naked singularity appears
along $m_{\ssc\Sigma}$. Another interesting situation, which appears in
\cite{cosmic,circle4}, is when the rotational symmetry only appears
after a conformal transformation of the boundary geometry. Of course,
the rotational symmetry is only a requirement of a specific CFT
analysis \cite{rt2,solo} and one can not expect that such symmetry
arises for a generic entangling surface. Hence, more generally, it
would be useful to develop a better understanding of the geometry of
the extremal bulk surface appearing in the calculation of holographic
EE, perhaps along the lines of \cite{broad}.

\acknowledgments

We would like to thank Horacio Casini, Matt Headrick, Ted Jacobson,
Miguel Paulos, Aninda Sinha and Brian Swingle for useful conversations
and correspondence. Further we would extend special thanks to Jan de
Boer, Manuela Kulaxizi and Andre Parnachev for useful discussions and
their patience in coordinating the posting of our respective papers.
Research at Perimeter Institute is supported by the Government of
Canada through Industry Canada and by the Province of Ontario through
the Ministry of Research \& Innovation. RCM also acknowledges support
from an NSERC Discovery grant and funding from the Canadian Institute
for Advanced Research.

%APPENDIX
\appendix

\section{Fefferman-Graham expansion}\label{schwimmer}

In a holographic framework, as long as the boundary field theory is
conformal in the UV, the dual geometry will approach AdS
asymptotically. In this context, the bulk spacetime will admit a
Fefferman-Graham expansion as follows \cite{feffer} (see also
\cite{sken,skenderis}):
 \be
ds^2 = \frac{\tilde{L}^2}{4}\frac{d\rho^2}{\rho^2} +
\frac{1}{\rho}\,g_{i j}(x,\rho)\,dx^i dx^j\,,
 \ee
where $g_{i j}$ admits a Taylor series expansion\footnote{The power
series expansion in $\rho$ can be altered in the back-reaction from
other nontrivial fields. A simple example would be when the boundary
CFT is deformed by a relevant operator --- \eg see
\cite{skenderis,universal}.} in the radial coordinate $\rho$
 \be
g_{i j}(x,\rho)= \gz_{i j}(x^i) + \rho \go_{i j}(x^i) + \rho^2 \gs_{i
j}(x^i) + \cdots\,.
 \labell{expand}
 \ee
The leading term $\overset{\scriptscriptstyle{(0)}}{g}_{ij}$ is
identified with the boundary CFT metric. The subsequent coefficients
$\overset{\scriptscriptstyle{(n)}}{g}_{ij}$ can be determined in terms
of $\overset{\scriptscriptstyle{(0)}}{g}_{ij}$ order by order by
expanding the gravitational equations of motion ---  although, for even
$d$, this expansion breaks down at order $n= d/2$  where an additional
logarithmic term appears.

It was shown in \cite{adam} that these coefficients
$\overset{\scriptscriptstyle{(n)}}{g}_{ij}$ are largely fixed by
conformal symmetries at the boundary, up to some small ambiguity that
must be fixed by the equations of motion. This procedure applies for
all $n<d/2$ for either odd or even $d$. Specifically,
$\overset{\scriptscriptstyle{(1)}}{g}_{ij}$ and
$\overset{\scriptscriptstyle{(2)}}{g}_{ij}$ for arbitrary
$\overset{\scriptscriptstyle{(0)}}{g}_{ij}$ are given by \cite{adam}
 \bea
\go_{i j} &=& -\frac{\tilde{L}^2}{d-2}\bigg(R_{i j}-
\frac{\overset{\scriptscriptstyle{(0)}}{g}_{ij}}{2(d-2)}R \bigg)\, \nonumber\\
\gs_{i j} &=& \tilde{L}^4\bigg(k_1\,  C_{m n k l}C^{m n k l}\gz_{i j} +
k_2\, C_{i k l m}C_{j}^{~~k l m} \nonumber \\
&& + \frac{1}{d-4}\bigg[\frac{1}{8(d-1)}\nabla_i\nabla_jR - \frac{1}{4(d-2)}\Box R_{i j}+
\frac{1}{8(d-1)(d-2)}\Box R \gz_{i j} \nonumber\\
&&-\frac{1}{2(d-2)}R^{k l}R_{i k j l} + \frac{d-4}{2(d-2)^2}R_{i}^{~k}R_{j k}+
\frac{1}{(d-1)(d-2)^2}RR_{i j}\nonumber \\
&&+\frac{1}{4(d-2)^2}R^{k l}R_{k l}\gz_{i j}
-\frac{3d}{16(d-1)^2(d-2)^2}R^2 \gz_{i j}\bigg]\bigg)\,,
 \label{metricexpand}
 \eea
where the curvature tensors are evaluated for the boundary metric
$\overset{\scriptscriptstyle{(0)}}{g}_{ij}$. The constant coefficients
$k_1$ and $k_2$ are precisely the remaining ambiguities (at this order)
which cannot be determined from symmetries alone. For the cubic
Lovelock gravity theory (\ref{GBact3}) in seven dimensions, they
are determined to be
 \be
k_1 = \frac{\fin \lambda  + 3\fin^2\mu}{160
(1-2\fin\lambda-3\fin^2\mu)},\quad k_2= -\frac{\fin\lambda + 3\fin^2\mu
}{24 (1-2\fin \lambda -3\fin^2\mu)}\,.
 \labell{ambig}
 \ee
These results \reef{metricexpand} are all that is needed to extract the
logarithmic divergent term in the holographic entanglement entropy for
the Lovelock theory in section \ref{EE6}.

\section{Curved boundaries} \label{curveb}

In the following, we consider the cubic Lovelock theory in seven (bulk)
dimensions, with the action in eq.~\reef{GBact3}. The (vacuum)
equations of motion are given by
 \bea
 R_{\mu\nu}+\frac{L^2}{6}\lambda\,\left(R_{\mu \sigma\rho\tau}R_\nu{}^{ \sigma\rho\tau}  - 2R_{\mu \rho}R_\nu{}^\rho
 -2R_{\mu\rho\nu\sigma}R^{\rho\sigma} + R\,R_{\mu\nu}\right) \quad&&
 \labell{motion}\\
 -{L^4 \over 8}\mu \Big( R_{\mu\nu}R^2-4R_{\mu\nu}R^{\rho\sigma}R_{\rho\sigma}+R_{\mu\nu}R_{\rho\sigma\tau\chi}
 R^{\rho\sigma\tau\chi}-4R_{\mu\rho\nu\sigma}R^{\rho\sigma}R
 +8R_{\mu\rho\nu\sigma}R^{\rho \chi \sigma \omega}R_{\chi \omega}
 &&
 \non
+8R_{\mu\rho\nu\sigma}R^{\rho
\tau}R^\sigma_\tau-4R_{\mu\rho\nu\sigma}R^\rho_{~\chi\tau\omega}R^{\sigma
\chi\tau\omega} -4R_{\mu\rho}R_\nu^\rho R +8 R_{\mu
\sigma\rho\chi}R^\rho_\nu R^{\sigma\chi}
  &&
 \non
+4R_{\mu \sigma\tau\omega}R^{\tau\omega\sigma}_{\quad
\rho}R^\rho_\nu+2R_{\mu
\sigma\tau\omega}R_\nu^{~\sigma\tau\omega}R-4R_{\mu
\sigma\tau\omega}R_{\nu\rho}^{\tau\omega}R^{\sigma\rho}+4R_{\nu \sigma
\tau\omega}R^{\tau\omega\sigma\rho}R_{\mu \rho}
 &&
 \non
+2R_{\mu \tau\rho\sigma}R_\nu^{~\tau
\chi\omega}R^{\rho\sigma}_{\chi\omega}+8R_{\mu \rho}R_{\nu
\sigma}R^{\rho\sigma}-8R_{\mu \sigma\tau\omega}R_{\nu~~
\rho}^{~\sigma\tau}R^{\omega\rho}
 +8R_{\nu \sigma\tau\omega}R_\mu^\tau R^{\sigma\omega}
 &&
 \non
 -8R_{\mu\sigma\tau\omega}R^{\tau \chi}{}_{\nu \rho}R^{\omega \rho \sigma}_{\quad \chi}\Big)
 -\frac{1}{2}g_{\mu\nu}\left(\frac{30}{L^2}+R+\frac{L^2}{12}\lambda\,\mathcal{L}_4-
 \frac{L^4}{24}\mu\,\mathcal{L}_6\right)
 &=&0\,.
\nonumber
 \eea
The above equations can be found in many places in the literature, \eg
see \cite{deg}.

In sections \ref{unbroken} and \ref{broken}, we wish to study
asymptotically AdS$_7$ solutions where the boundary metric is not
conformally flat. The simplest approach is to construct these solutions
using the Fefferman-Graham expansion near the AdS boundary, as in
appendix \ref{schwimmer}. In eq.~\reef{metricexpand}, we provide
explicit formulae for $\overset{\scriptscriptstyle{(1)}}{g}_{ij}$ and
$\overset{\scriptscriptstyle{(2)}}{g}_{ij}$, constructed for a given
boundary metric $\overset{\scriptscriptstyle{(0)}}{g}_{ij}$. Instead we
produced our results here by explicitly solving the equations of motion
\reef{motion}, order by order in the asymptotic expansion. For the
examples considered in sections \ref{unbroken} and \ref{broken}, we
found:

\vskip1em \noindent{\bf a) $R\times S^2\times S^3$:}

Consider the following metric ansatz:
\begin{equation}
ds^2=\frac{\tilde L^2}{z^2}\left(dz^2+f_1(z)\,dt^2+f_2(z)\,\r22\, d\Omega^2_2
+f_3(z)\,\r12\, d\Omega^2_3\right)
\,,
\labell{metric}
\end{equation}
where $d\Omega^2_2$ and $d\Omega^2_3$ are standard round metrics on a
unit two-sphere and three-sphere, respectively. We expand around the
asymptotic boundary with
 \beq
 f_i(z)=1+\sum_{j=1}^\infty k_{i,j}\, z^{2j} \,.
 \labell{expansion}
 \eeq
Now aided by the appropriate computer software, we solve the equations
of motion \reef{motion} order by order in our expansion in powers of
$z^2$. To leading order, we find the familiar expression
 \be
 1-\fin+\lambda\,\fin^2+\mu\,\fin^3=0\,.
 \labell{const0}
 \ee
At second order, we find:
 \be
 k_{1,1} = \frac{3\,\r12+\r22}{20\,\r12\,\r22}\,,\quad
 k_{2,1} = -\frac{4\,\r12-3\,\r22}{20\,\r12\,\r22}\,,\quad
 k_{3,1} = \frac{\r12-7\,\r22}{20\,\r22\,\r32}\,.
 \labell{const1}
 \ee
At the next order, the coefficients can be expressed as:
 \bea
 k_{1,2} &=& \frac{2\r14(8-9\fin\lambda-3\fin^2\mu)
 -2\r12\r22(27-86\fin\lambda-177\fin^2\mu)+\r24(69-142\fin\lambda-219\fin^2\mu)
 }{1600\,(1-2\fin\lambda-3\fin^2\,\mu)\,\r14\,\r24}\,,\quad
 \nonumber\\
 k_{2,2} &=& \frac{\r24(69-142\fin\lambda-219\fin^2\mu)+6\r12\r22(1-8\fin\lambda-21\fin^2\mu)
 -2\r14(7+4\fin\lambda+33\fin^2\mu)}{1600\,(1-2\fin\lambda-3\fin^2\,\mu)\,\r14\,\r24}\,,\quad
 \labell{const2}\\
 k_{3,2} &=& \frac{6\r14(8-9\fin\lambda-3\fin^2\mu)-2\r12\r22(21-38\fin\lambda-51\fin^2\mu)
 -\r24(33-54\fin\lambda-63\fin^2\mu)}{4800\,(1-2\fin\lambda-3\fin^2\,\mu)\,\r14\,\r24}\,.
 \nonumber
 \eea

\vskip1em\noindent{\bf b) $R^3\times S^3$:}

Consider the following metric ansatz:
\begin{equation}
ds^2=\frac{L^2}{z^2}\left(dz^2+f_1(z)\,\left(dt^2+dx^2
+dy^2\right)
+f_3(z)\,\r12\, d\Omega^2_3\right)
\,,
\labell{metric3}
\end{equation}
where $d\Omega^2_3$ is the standard round metric on a unit
three-sphere. We expand around the asymptotic boundary with the same
expressions as in eq.~\reef{expansion} and solve the equations
\reef{motion} order by order in our expansion in powers of $z^2$. As
expected, to leading order, we again recover eq.~\reef{const0}. At
second order, we find:
 \be
 k_{1,1} = \frac{3}{20\,\r12} \,,\quad k_{3,1} = -\frac{7}{20\,\r12} \,.
 \labell{const13}
 \ee
At the next order, the coefficients can be expressed as:
 \be
 k_{1,2} = \frac{-69+142\fin\lambda+219\fin^2\mu}{1600(-1+2\fin\lambda+3\fin^2\mu)\,\r14}
 \,,\quad k_{3,2}= \frac{11-18\fin\lambda-21\fin^2\mu}{1600(-1+2\fin\lambda+3\fin^2\mu),\r14} \,.
 \labell{const23}
 \ee

\vskip1em\noindent{\bf c) $R^2\times S^4$:}

Consider the following metric ansatz:
\begin{equation}
ds^2=\frac{L^2}{z^2}\left(dz^2+f_1(z)\,\(dt^2+dx^2\)
+f_2(z)\,\r12\, d\Omega^2_4\right)
\,,
\labell{metric2}
\end{equation}
where $d\Omega^2_4$ is the standard round metric on a unit four-sphere.
We proceed as above solving the equations \reef{motion} order by order.
To leading order, we again recover eq.~\reef{const0}. At second order,
we find:
 \be
 k_{1,1} = \frac{3}{10\,\r12}\,,\quad k_{2,1} = -\frac{9}{20\,\r12} \,.
 \labell{const12}
 \ee
At the next order, the coefficients can be expressed as:
 \be
 k_{1,2} = \frac{9(-7+16\fin\lambda+27\fin^2\mu)}{800(-1+2\fin\lambda+3\fin^2\mu)\,\r14}
 \,,\quad k_{2,2}= \frac{-18+29\fin\lambda+33\fin^2\mu}{800(-1+2\fin\lambda+3\fin^2\mu)\,\r14} \,.
 \labell{const22}
 \ee

\vskip1em\noindent{\bf d) $ S^3\times S^3$:}

Consider the following metric ansatz:
\begin{equation}
ds^2=\frac{L^2}{z^2}\left(dz^2+f_1(z)\,\r12\, d\Omega^2_3
+f_2(z)\,\r22\, d\Omega^2_3\right)
\,,
\end{equation}
where $d\Omega^2_3$ is the standard round metric on a unit
three-sphere. Proceeding as above yields eq.~\reef{const0} to leading
order, whereas at second order, we obtain
 \be
 k_{1,1} = \frac{3\r12-7\r22}{20\,\r12\r22}\,,
 \quad k_{2,1} =  \frac{3\r22-7\r12}{20\,\r12\r22} \,.
 \ee
At third order, we have
 \bea
 k_{1,2} &=& \frac{\r24(11-18\fin\lambda-21\fin^2\mu)+2\r12\r22(21-38\fin\lambda-51\fin^2\mu)
 -\r14(69-142\fin\lambda-219\fin^2\mu)}{1600\,(-1+2\fin\lambda+3\fin^2\,\mu)\,\r14\,\r24}\,,\quad
 \nonumber\\
 k_{2,2} &=& \frac{\r14(11-18\fin\lambda-21\fin^2\mu)+2\r12\r22(21-38\fin\lambda-51\fin^2\mu)
 -\r24(69-142\fin\lambda-219\fin^2\mu)}{1600\,(-1+2\fin\lambda+3\fin^2\,\mu)\,\r14\,\r24}\,.\quad
 \nonumber
 \eea

\vskip1em Note that if one trades, \eg $S^2\times S^3$ for $H^2\times
H^3$, the signs of all the curvatures are reversed. Hence the
contributions to the six-dimensional trace anomaly in the boundary
theory are simply flipped and so we do not expect to get a distinct
test of our holographic entanglement entropy for Lovelock gravity.
Other simple boundary manifolds which should give distinct results for
the trace anomaly include: $R^4\times S^2$,
$R^2\times\left(S^2\right)^2$, $R^2\times S^2\times H^2$.

\section{EE in the GB gravity}
\label{appx:EE-disk}

In this appendix we consider a $d$-dimensional boundary CFT dual to a
GB gravity. The EE is investigated in the case when the entangling
surface is either a sphere or a cylinder. Thus the number of terms in
eq.~(\ref{GBactg}) is restricted to $p_{max}=2$
 \be
I=\frac {1}{2\lp^{d-1}}\int d^{d+1}x \sqrt{-g}
\Big[\frac{d(d-1)}{L^2}+R+\frac{{L}^2\,\lambda}{(d-2)(d-3)}\mathcal{L}_4
%\\
%+{\frac{{L}^4\,\mu}{(d-2)(d-3)(d-4)(d-5)}\mathcal{L}_6}
\Big] \,,
 \ee
and hence eq.~(\ref{Waldformula3}) reduces to
 \be
 S_{\mt{JM}}=\frac {2\pi}{\lp^{d-1}}\int_{m} d^{d-1}x\sqrt{h}
 \[ 1+\frac{2\,L^2\,\lambda}{(d-2)(d-3)}\,R \]
 %+\frac{3L^4\,\mu}{(d-2)(d-3)(d-4)(d-5)}\,\mathcal{L}_4\]
 +\frac{8\pi}{\lp^{d-1}}\frac{\,L^2\,\lambda}{(d-2)(d-3)}\int_{\del m}\mathcal{K}
 \labell{eqn:EEgendim}
\ee
%{\bf [Extra surface term should be included. arXiv:0807.1256
%[hep-th] ]}

The $AdS_{d+1}$ metric
 \be
 ds^2=\frac{\tilde L^2}{z^{2}}(dz^2-dt^2+\sum_{i=1}^{d-1}dx_i^2)
 \quad ,
 \ee
is an exact solution of the equations of motion in the GB gravity. We
introduce a short distance cut-off in the boundary CFT here by setting
a minimum value of $z$: $z=z_{UV}= \delta$. In what follows, we choose
either $\sum_i dx_i^2=dr^2+r^2d\Omega^2_{d-2}$ or $\sum_i
dx_i^2=dv^2+dr^2+r^2d\Omega^2_{d-3}$ when $V$ corresponds to a ball,
$A_D:=\{x_i|r\leq R\}$ or a solid cylinder, $A_C=\{x_i|r\leq R\}$,
respectively. Hence, the induced metric on the static minimal surface
in the $AdS_{d+1}$ bounded by either $\partial A_D$ or $\partial A_C$,
is given respectively by
 \be
 h_{a b}^Ddx^{a}dx^{b}=\frac{\tilde L^2}{z^{2}}[(\dot r^2+\dot z^2)du^2+r^2d\Omega^2_{d-2}]
 \quad,
  \label{eqn:ind-metric-disk}
 \ee
and
 \be
 h_{a b}^Cdx^{a}dx^{b}=\frac{\tilde L^2}{z^{2}}[(\dot r^2+\dot z^2)du^2+dv^2+r^2d\Omega^2_{d-3}]
 \quad,
 \label{eqn:ind-metric-cylinder}
 \ee
where $u$ parametrizes the minimal surface in the $(z,r)$ plane and
dot denotes the derivative with respect to $u$. Both expressions are
of the form
 \bea
 ds^2&=&ds^2_X+\sum_ie^{2F_i}ds_{Y_i}^2 %\Rightarrow
 \label{eqn:fibration-formula}
 \eea
where the conformal factors depend on the $x$ coordinates only,
$F_i=F_i(x)$. In this case, one can conveniently decompose the
curvature tensor and the associated scalars in terms of $F_i$ fields
and the curvature tensor of the $X$ space. The related useful formulae
are summarized in appendix \ref{appx:fibration}.

\subsection{EE for a sphere with general $d$}

In this case (comparing eqs.~(\ref{eqn:ind-metric-disk}) and
\reef{eqn:fibration-formula}) we identify a one-dimensional space along
the $u$-direction with $X$ of (\ref{eqn:fibration-formula}) and the
$(d-2)$-dimensional sphere with radius $\tilde L$ is identified with
$Y$, whereas
 \bea
 e^{2F}&=&{ r^2 \over z^2} \quad\Rightarrow\quad F=\ln(r/z)~.
 \eea
In particular,
 \be
 R_{acbd}^Y={1 \over \tilde L^2}(g_{ab}^Yg_{cd}^Y-g_{ad}^Yg_{bc}^Y)
 \quad\Rightarrow\quad R_{ab}^Y={d-3 \over \tilde L^2}g_{ab}^Y~, \quad R^Y={(d-2)(d-3) \over \tilde L^2}
 \quad,
 \ee
where $g_{ab}^Y$ is the metric of the unit $(d-2)$-dimensional
sphere. Using (\ref{fibrcurv}) yields
 \be
 R_D=\,e^{-2F}\,{(d-2)(d-3) \over \tilde L^2}-2(d-2)\Delta_X(F)-(d-2)(d-1)h^{uu} \dot F^2 ~,
  \label{eqn:disk-ricci-scalar}
 \ee
where  as before  `dot'denotes the derivative with respect to $u$, and
 \be
 h_{uu}=(h^{uu})^{-1}={\tilde L^2 \over z^{2}}(\dot r^2+\dot z^2)
 \quad\Rightarrow\quad \Delta_X(F)={1 \over \sqrt{h_{uu}} }\,\del_u\,(\sqrt{h_{uu}}\,h^{uu}\dot F)~.
 \ee

Let us evaluate now the extrinsic curvature $\mathcal{K}$. The normal
outward unit vector to the boundary surface defined by $u=u_i$, or
equivalently by $r(u_i)=R, z(u_i)=\de$, is given by
 \be
 n_a=-\sqrt{h_{uu}}\,\delta_{u a}
 \quad ,
 \ee
where $a$ runs over $u$ and a $(d-2)$-dimensional sphere, thus
($i,k$ below run over the $(d-2)$-dimensional sphere only)
 \be
 \mathcal{K}=h^{a b}\nabla_a n_b |_{u=u_i}=\[h^{uu}\nabla_u n_u+g^{ik}\nabla_i n_k\]_{u=u_i}
 =-{(d-2)\dot F\over \sqrt{h_{uu}} }\,e^{(d-2)F}\Big|_{u=u_i}
 ~,
 \ee
where we used $2(\det h_{ab})^{-1/2}\del_u \sqrt{\det
h_{ab}}=h^{ab}\del_u h_{ab}$. As a result, we get
 \be
 \int_{\del m}\mathcal{K} =-\tilde L^{d-2}S_{d-2}\, h_{uu}^{-1/2}\,\del_u\,e^{(d-2)F}\Big|_{u_i}
 \quad .
 \label{eqn:extrinsic-disk}
 \ee

Having these results at hand, one can show that the minimal surface of (\ref{eqn:EEgendim}),
can be conveniently parameterized as follows
 \be
 r(u)=R\cos(u/R)~ , \quad z(u)=R\sin (u/R)~, \quad{\rm where}\ \
  \de\leq u \leq {\pi \over 2}R~.
 \label{eqn:disk-parametrization}
 \ee
Here we need to introduce the ratio of $u$ to some scale in the
argument of the trigonometric functions above in order to maintain the
correct dimensions. We chose $R$ as the natural scale, however, any other scale can be used instead.
Let us proceed by substituting (\ref{eqn:disk-ricci-scalar}) into (\ref{eqn:EEgendim}) and integrating
by parts
 \begin{align}
 S_{\mt{JM}}=&-\frac {8\pi\,L^{d-2}S_{d-2}\,\lambda}{\lp^{d-1}(d-2)(d-3)}
 \,\tilde{L}^2\,\[ h_{uu}^{-1/2}\del_u\,e^{(d-2)F}\]_{u_i}^{u_f}
 \non
 &+\frac {2\pi\,L^{d-2}S_{d-2}}{\lp^{d-1}} \int du \sqrt{h_{uu}}
 \,e^{(d-2)F}\Big(1 +2\,(\tilde{L}/L)^2\,\lambda \Big[ e^{-2F}
 +\,L^2h^{uu}\dot F^2\Big]\Big)
 \non
 &+\frac{8\pi}{\lp^{d-1}}\frac{\,\tilde{L}^2\,\lambda }{(d-2)(d-3)}\int_{\del m}\mathcal{K}
 ~.
 \end{align}
According to (\ref{eqn:extrinsic-disk}) the Gibbons-Hawking term
precisely cancels the boundary contribution which corresponds to the
lower bound of the first term in the above expression (the upper
bound vanishes due to symmetry). Substituting now the general
parametrization
 \be
r(u)=f(u/R)\cos(u/R)~ , \quad z(u)=f(u/R)\sin (u/R)~, \quad \de\leq u
\leq {\pi \over 2}R~,
 \ee
yields (with $x=u/R$)
 \begin{multline}
S_{\mt{JM}}=\frac {2\pi\,L^{d-1}S_{d-2}}{\lp^{d-1}} \int_{\de/
R}^{\pi \over 2} dx \, {\cos(x)^{d-2} \over \sin(x)^{\,
d-1}}\,\sqrt{1+\Big({d\ln\,f\over dx}\Big)^2}
 \\
\times\Big(1 +2\,(\tilde{L}/L)^2\,\lambda\Big[
\tan^2(x)+{\cos^{-2}(x)\over 1+\Big({d\ln\,f\over dx}\Big)^2}\Big]\Big)
 \end{multline}
Since the integrand depends on `time' $x$ and on the square of the
`velocity' $v(x):=d\ln\,f/dx$, the corresponding Euler-Lagrange
equation is
 \be
 {d \over dx} \( {\partial L(x,v^2(x)) \over \partial v(x)} \)=0
 \quad ,
 \ee
and it admits the solution $v=0 \Leftrightarrow f=const$. Plugging
(\ref{eqn:disk-parametrization}) into (\ref{eqn:disk-ricci-scalar}),
yields
 \bea
 R_D&=&-{(d-1)(d-2) \over \tilde L^2}~.
 \labell{eqn:disk-ricci-scalar1}
 \eea

Substituting this result back into (\ref{eqn:EEgendim}), we finally obtain
 \be
 S_{\mt{JM}}=\frac {2 \pi}{\lp^{d-1}}\[1
 -2\,\frac{d-1}{d-3}\,\fin\,\lambda
 \]\int_{m} d^{d-1}x\sqrt{\det h^D_{ab}} + \frac{8\pi}{\lp^{d-1}}\frac{\,\tilde{L}^2\,\lambda }{(d-2)(d-3)}\int_{\del m}\mathcal{K}
 ~,
 \ee
where the area of the surface is given by
 \be
 \int_{m} d^{d-1}x\sqrt{\det h^D_{ab}}
 =\tilde L^{d-1}S_{d-2}\int_{\delta/R}^{1}dy \, {(1-y^2)^{d-3 \over 2} \over y^{\, d-1}}
 =(-)^{d/2}\tilde L^{d-1}{~2 \pi^{d/2-1} \over \Gamma(d/2)}\log (\delta/R)+\cdots
 \labell{diskarea}
 \ee
where $S_{d-2}$ is the area of the unit ($d-2$)-dimensional sphere and
in the last equality we assumed that $d$ is even, since only in that
case does the integral contains a subleading logarithmic divergence,
which can be evaluated by expanding the integrand in powers of $y$. The
rest of the terms are encoded in ellipsis. As a result, the universal
term (for even $d$) in the EE is given by
 \be
  S_{\mt{JM}}=(-)^{d/2-1}  {4\pi^{d/2} \over \Gamma(d/2)}
\frac {\tilde
L^{d-1}}{\lp^{d-1}}\[1-2\,\frac{d-1}{d-3}\,\fin\,\lambda\]\log
(l/\delta)+\cdots\,.
  \labell{eqn:EEdisk}
 \ee
Now comparing this result with eq.~\reef{llA}, we recognize that the
pre-factor is proportional to $A$. In fact, our result here matches
that in \cite{circle4}.

\subsection{Spherical entangling surfaces beyond GB gravity.}

It was shown in \cite{circle4} that when the entangling surface
corresponds to a sphere, the universal term in the EE will be always
proportional to the A-type anomaly for even $d$. Therefore inclusion of
the higher order interactions (\ref{term0}) in the Lovelock gravity
will reconstruct (\ref{llA}) term by term. To illustrate this idea, let
us do one step beyond the GB interaction by taking $p_{max}=3$ in
\reef{GBactg}
\begin{multline}
I=\frac {1}{2\lp^{d-1}}\int d^{d+1}x \sqrt{-g}
\Big[\frac{d(d-1)}{L^2}+R+\frac{{L}^2\,\lambda}{(d-2)(d-3)}\mathcal{L}_4
\\
+{\frac{{L}^4\,\mu}{(d-2)(d-3)(d-4)(d-5)}\mathcal{L}_6} \Big]+\ldots \,,
\end {multline}
then \reef{Waldformula3} becomes
\be
 S_{\mt{JM}}=\frac {2\pi}{\lp^{d-1}}\int_{m} d^{d-1}x\sqrt{h}
 \[ 1+\frac{2\,L^2\,\lambda}{(d-2)(d-3)}\,\R
 +\frac{3L^4\,\mu}{(d-2)(d-3)(d-4)(d-5)}\,\mathcal{L}_4(\R)\]+\ldots
 %+\frac{8\pi}{\lp^{d-1}}\frac{\,L^2\,\lambda}{(d-2)(d-3)}\int_{\del m}\mathcal{K}
 \labell{eqn:EEgendim2}
 \ee
The ellipsis denotes the surface terms \cite{surf}, which are
suppressed since they do not contribute to the universal divergence
explored in what follows.

Using (\ref{eqn:curvature-inv}) one finds
 \begin{align}
 R_{D\,\mu\nu\rho\sigma}R^{\mu\nu\rho\sigma}_D=&{2\over (d-2)(d-3)}\[ e^{-2F}R^Y-(d-2)(d-3)h^{uu} \dot F^2 \]^2
 \non
 &\quad\quad\quad\quad\quad\quad\quad\quad\quad\quad\quad\quad\quad\quad\quad\quad
 +4(d-2)\[\Delta_X\,F+h^{uu} \dot F^2\]^2~,
 \non
 R_{D\,\mu\nu}R^{\mu\nu}_D=&{1\over d-2}\[ e^{-2F}R^Y-(d-2)(\Delta_X\,F+(d-2)h^{uu} \dot F^2) \]^2
 \non
 &\quad\quad\quad\quad\quad\quad\quad\quad\quad\quad\quad\quad\quad\quad\quad\quad
 +(d-2)^2\[\Delta_X\,F+h^{uu} \dot F^2\]^2~,
\end{align}
Substituting (\ref{eqn:disk-parametrization}) into these expressions,
yields\footnote{One can extend the argument presented in the case of GB
gravity to prove that \reef{eqn:disk-parametrization} minimizes
\reef{eqn:EEgendim2}. }
 \bea
 R_{D\,\mu\nu\rho\sigma}R^{\mu\nu\rho\sigma}_D&=&{2(d-1)(d-2) \over \tilde L^4}~,
 \non
 R_{D\,\mu\nu}R^{\mu\nu}_D&=&{(d-1)(d-2)^2 \over \tilde L^4}~.
\eea
Now combining the latter with (\ref{eqn:disk-ricci-scalar1}), we get
\begin{equation}
 \mathcal{L}_4=R_{D\,\mu\nu\rho\sigma}R^{\mu\nu\rho\sigma}_D-4 R_{D\,\mu\nu}R^{\mu\nu}_D+R_D^2={(d-1)(d-2)(d-3)(d-4)\over \tilde L^4}~.
\end{equation}
Plugging all the above into (\ref{eqn:EEgendim2}), we finally obtain
 \be
 S_{\mt{JM}}=\frac {2 \pi}{\lp^{d-1}}\[1
 -2\,\frac{d-1}{d-3}\,\fin\,\lambda+3\,\frac{d-1}{d-5}\,\fin^2\,\mu
 \]\int_{m} d^{d-1}x\sqrt{\det h^D_{ab}} + \ldots
 %\frac{8\pi}{\lp^{d-1}}\frac{\,\tilde{L}^2\,\lambda }{(d-2)(d-3)}\int_{\del m}\mathcal{K}
 \ee
Substituting \reef{diskarea} we recover \reef{Waldformula9} where $A$
is given by \reef{llA} with $p_{max}=3$. In fact again, this result for
the cubic Lovelock theory matches precisely with the expression derived
in \cite{circle4} for an arbitrary higher curvature theory.

\subsection{EE for a cylinder with general $d$} \label{cylinder99x}

In the case of (\ref{eqn:ind-metric-cylinder}), we identify a one
dimensional space along $u$-direction with $X$ of
(\ref{eqn:fibration-formula}), whereas a one dimensional space along
$v$-direction and a $(d-3)$-dimensional sphere with radius $L$ are
identified with $Y_1$ and $Y_2$ respectively. Hence,
 \bea
 e^{2F_1}&=&{\tilde L^2 \over z^2} \quad\Rightarrow\quad F_1=\ln(\tilde L/z)~,
 \non
 e^{2F_2}&=&{ r^2 \over z^2} \quad\Rightarrow\quad F_2=\ln(r/z)~.
 \eea
Substituting into (\ref{eqn:EEgendim}), yields
 \begin{multline}
 R_C=e^{-2F_2}\,{(d-3)(d-4) \over \tilde L^2}-2\Delta_X(F_1)-2(d-3)\Delta_X(F_2)-2h^{uu} \dot F_1^2
 \\-(d-3)(d-2)h^{uu} \dot F_2^2
 -2(d-3)h^{uu} \dot F_1\dot F_2~,
 \label{eqn:cylinder-ricci-scalar}
 \end{multline}
where dot denotes derivative with respect to $u$ and
 \be
 h_{uu}=(h^{uu})^{-1}={\tilde L^2 \over z^{2}}(\dot r^2+\dot z^2) \quad\Rightarrow\quad \Delta_X(F_i)={1 \over \sqrt{h_{uu}} }\,\del_u\,(\sqrt{h_{uu}}\,h^{uu}\dot F_i)~.
 \ee

Next we evaluate the extrinsic curvature $\mathcal{K}$. The normal
outward unit vector to the boundary surface defined by $u=u_i$, or
equivalently by $r(u_i)=R, z(u_i)=\delta$, is given by
 \be
 n_a=-\sqrt{h_{uu}}\,\delta_{ua}
 \quad ,
 \ee
where $a$ runs over $u$, $v$ and a $(d-3)$-dimensional sphere,
thus ($i,k$ below run over the $(d-3)$-dimensional sphere only)
 \be
 \mathcal{K}=g^{a b}\nabla_a n_b|_{u=u_i}=\[h^{uu}\nabla_u n_u+g^{vv}\nabla_v n_v+g^{ik}\nabla_i n_k\]_{u=u_i}
 =-{\,\del_u\,e^{F_1+(d-3)F_2}\over \sqrt{h_{uu}} }\Big|_{u=u_i}
 ~,
 \ee
where we used $2(\det h_{ab})^{-1/2}\del_u \sqrt{\det
h_{ab}}=h^{ab}\del_u h_{ab}$. Thus
 \be
 \int_{\del m}\mathcal{K} =-\tilde L^{d-3}S_{d-3}H\, h_{uu}^{-1/2}\,\del_u\,e^{F_1+(d-3)F_2}\Big|_{u_i}
 \quad ,
 \label{eqn:extrinsic-cylinder}
 \ee
where $H=\int\,dv$ is the length of the cylinder.

Substituting (\ref{eqn:cylinder-ricci-scalar}) into
(\ref{eqn:EEgendim}) and integrating by parts, yields
 \begin{align}
 S=-\frac {8\pi\,\tilde L^{d-3}S_{d-3}H\,\lambda}{\lp^{d-1}(d-2)(d-3)}\,L^2\,\[ h_{uu}^{-1/2}\del_u\,e^{F_1+(d-3)F_2}\]_{u_i}^{u_f}
 +\frac {2\pi\,\tilde L^{d-3}S_{d-3}H}{\lp^{d-1}} \int du \sqrt{h_{uu}}
 \non
 \times\,e^{F_1+(d-3)F_2}\Big(1 +\frac{2\,(L/\tilde L)^2\,\lambda}{(d-2)}\Big[ (d-4)e^{-2F_2}
 +2h^{uu}\,\tilde L^2\dot F_1\dot F_2 + (d-4)\,\tilde L^2h^{uu}\dot F_2^2\Big]\Big)
 \non
 +\frac{8\pi}{\lp^{d-1}}\frac{\,\tilde L^2\,\lambda}{(d-2)(d-3)}\int_{\del m}\mathcal{K}
 ~.
 \label{eqn:cylinder-gen-dim}
 \end{align}
According to (\ref{eqn:extrinsic-cylinder}), the Gibbons-Hawking term
cancels the lower bound of the first term in the above expression. In
contrast to the case of the ball, we did not succeed to find a closed
analytic expression for the surface which minimizes
\reef{eqn:cylinder-gen-dim} in general $d$. However, to evaluate the
universal divergence, one needs to know the asymptotic expansion of
such a surface to order which depends on $d$. Therefore to proceed
further, we must pick a particular value, \eg $d=4$, for the dimension
of the boundary CFT. We illustrate such computation in section
\ref{sec:EEcyl}.

\section{Curvature tensor for a warped geometry}
\label{appx:fibration}

In this appendix, we derive some results which are useful to evaluate
$S_\mt{JM}$ in appendix \ref{appx:EE-disk}. In particular, we determine
the general expression for the Riemann curvature tenor
$R_{\alpha\beta\gamma\delta}$, Ricci tensor $R_{\beta\delta}$ and Ricci
scalar $R$ for a warped geometry of the form:
 \begin{eqnarray}
 ds^2&=&ds^2_X+\sum_ie^{2F_i}ds_{Y_i}^2 %\Rightarrow
 \end{eqnarray}
where the conformal factors depend on the $x$ coordinates in the base
$X$, \ie $F_i=F_i(x)$. Our convention for the curvature (which matches
\cite{mtw}) is
 \begin{equation}
R_{\alpha\beta\gamma\delta}={1 \over 2}(g_{\alpha\delta,\gamma\beta}
+g_{\beta\gamma,\alpha\delta}-g_{\alpha\gamma,\beta\delta}-g_{\beta\delta,\alpha\gamma})
+g_{\mu\nu}(\Gamma^\mu{}_{\beta\gamma}\Gamma^\nu,{}_{\alpha\delta}
-\Gamma^\mu{}_{\beta\delta}\Gamma^\nu{}_{\alpha\gamma})~.
 \end{equation}
In what follows Greek letters $\alpha,\beta,\gamma,...$ run over the
base space $X$, whereas Greek letters with a subscript
$\alpha_i,\beta_i,\gamma_i,...$ run over directions in the fibre spaces
$Y_i$. The nonvanishing components of the Christoffel symbol are given
by
 \begin{equation}
 \Gamma_{\alpha,\beta\gamma}=\Gamma_{\alpha,\beta\gamma}^{X},~
 \Gamma_{\alpha_i,\beta_i\gamma_i}=e^{2F}\Gamma_{\alpha_i,\beta_i\gamma_i}^{Y_i},~
 \Gamma_{\alpha_i,\beta_i\gamma}=\partial_\gamma F_i \,e^{2F_i}g_{\alpha_i\beta_i}^{Y_i},~
 \Gamma_{\alpha,\beta_i\gamma_i}=-\partial_\alpha F_i \,
 e^{2F_i}g_{\beta_i\gamma_i}^{Y_i},
 \end{equation}
where $\Gamma_{\alpha,\beta\gamma}=g_{\al\de}
\,\Gamma^\delta{}_{\beta\gamma}$. Further we introduced a notation
where superscript $X$ or $Y_i$ indicates that the corresponding
quantity, above the Christoffel symbol, is calculated for the metric on
the corresponding space. It follows that the non-vanishing components
of the curvature tensor are
 \begin{eqnarray}
 R_{\alpha\beta\gamma\delta}&=&R_{\alpha\beta\gamma\delta}^X~,
 \nonumber \\
 R_{\alpha\beta_i\gamma\delta_i}&=&-e^{2F_i}\left( \nabla_{\alpha}\nabla_\gamma
 F_i+\partial_\alpha F_i\,\partial_\gamma F_i \right)g_{\beta_i\delta_i}^{Y_i}
 ~,
 \nonumber \\
 R_{\alpha_i\beta_i\gamma_i\delta_i}&=&e^{2F_i}R_{\alpha_i\beta_i\gamma_i\delta_i}^{Y_i}
 +(\partial F_i)^2 \, e^{4F_i} \(g_{\beta_i\gamma_i}^{Y_i}g_{\alpha_i\delta_i}^{Y_i}
 -g_{\beta_i\delta_i}^{Y_i}g_{\alpha_i\gamma_i}^{Y_i}\)~,
 \nonumber \\
 R_{\alpha_i\beta_j\gamma_i\delta_j}&=&-(\partial F_i\cdot\partial F_j)
 \,e^{2(F_i+F_j)}g_{\beta_j\delta_j}^{Y_j}g_{\alpha_i\gamma_i}^{Y_i}
 \quad \mathrm{with}\quad i\neq j ~,
 \end{eqnarray}
where all derivatives are evaluated in the $X$ space and $\nabla$
denotes the covariant derivative compatible with the metric on $X$.
Thus the non-vanishing components of the Ricci tensor are
 \begin{eqnarray}
 R_{\beta\delta}&=&R_{\beta\delta}^X-\sum_i d_i \( \nabla_{\beta} \nabla_{\delta}F_i
 +\partial_\beta F_i\,\partial_\delta F_i\)~,
 \nonumber \\
 R_{\beta_i\delta_i}&=&R_{\beta_i\delta_i}^{Y_i}-(\nabla^2 F_i)\,
e^{2F_i}g_{\beta_i\delta_i}^{Y_i}
 -e^{2F_i}g_{\beta_i\delta_i}^{Y_i}\sum_{j}d_j(\partial F_i\cdot\partial F_j)~.
 \end{eqnarray}
where $d_i$ corresponds to the dimension of the space $Y_i$ and the
Laplacian $\nabla^2$ is again evaluated on $X$. Finally, the Ricci
scalar is
 \begin{eqnarray}
R&=&R^X+\sum_i\[e^{-2F_i}R^{Y_i}-2d_i(\nabla^2 F_i)-d_i
(\partial F_i)^2\]-\sum_{ij}d_id_j(\partial F_i\cdot\partial F_j)
 \,,
 \labell{fibrcurv}
 \end{eqnarray}
where $R^X$, $R^{Y_i}$ are the Ricci scalars of the spaces $X$ and
$Y_i$, respectively. In particular, using these expressions, one finds
the following following expressions for various invariants
 \bea
 R_{abcd}R^{abcd}&=&R_{\alpha\beta\gamma\delta}^X\, R^{X\alpha\beta\gamma\delta}
 +\sum_i\Big[ e^{-4F_i}R_{\alpha_i\beta_i\gamma_i\delta_i}^{Y_i}
 R^{Y_i\,\alpha_i\beta_i\gamma_i\delta_i} -4e^{-2F_i}R^{Y_i}(\partial F_i)^2
 \nonumber \\
 &&+4\,d_i\nabla_\alpha\nabla_\gamma F_i\nabla^\alpha\nabla^\gamma F_i
 +8d_i\nabla_\alpha\nabla_\gamma F_i\,\partial^\alpha F_i\,\partial^\gamma F_i-2d_i(d_i-1)(\partial F_i
 \cdot\partial F_i)^2\Big]
 \nonumber \\
 &&+4\sum_{ij}d_id_j(\partial F_i\cdot\partial F_j)^2~,
 \nonumber \\
 R_{ab}R^{ab}&=&R_{\beta\delta}^X\, R^{X\beta\delta}
 +\sum_i\Big[ e^{-4F_i}R_{\beta_i\delta_i}^{Y_i} R^{Y_i\,\beta_i\delta_i}-2d_i R_{\beta\delta}^X
 (\nabla^\bt\nabla^\de F_i+\partial^\beta F_i\,\partial^\delta F_i)
 \nonumber \\
 &&\qquad\qquad\qquad\qquad+d_i(\nabla^2\,F)^2-2\, e^{-2F_i}R^{Y_i}\nabla^2\,F_i \Big]
 \nonumber \\
 &&+\sum_{ij}\Big[ d_id_j(\nabla_\beta\nabla_\delta F_i+\partial_\beta F_i\,\partial_\delta F_i)
 (\nabla^\beta\nabla^\delta F_j+\partial^\beta F_j\,\partial^\delta F_j)
 \label{eqn:curvature-inv} \\
 &&-2d_j(e^{-2F_i}R^{Y_i}-d_i\nabla^2\,F_i)(\partial F_i\cdot\partial F_j) \Big]
 +\sum_{i,j,k}d_id_jd_k(\partial F_i\cdot\partial F_j)(\partial F_i\cdot\partial F_k)
 \nonumber
 \eea

\end{document}